# Probing the regoliths of the classical Uranian satellites: Are their surfaces mantled by a layer of tiny $H_2O$ ice grains?

Richard J. Cartwright[a,b,1], Joshua P. Emery[c,1], William M. Grundy[d], Dale P. Cruikshank[b], Chloe B. Beddingfield[a,b], Noemi Pinilla-Alonso[e]

[a]*Carl Sagan Center, SETI Institute;* [b]*NASA Ames Research Center;* [c]*Northern Arizona University;* [d]*Lowell Observatory;* [e]*University of Central Florida, Florida Space Institute*

## Abstract

We investigate whether the surfaces of the classical moons of Uranus are compositionally stratified, with a thin veneer of mostly tiny $H_2O$ ice grains (≤ 2 µm diameters) mantling a lower layer composed of larger grains of $H_2O$ ice, dark material, and $CO_2$ ice (~10 – 50 µm diameters). Near-infrared observations (~1 – 2.5 µm) have determined that the $H_2O$ ice-rich surfaces of these moons are overprinted by concentrated deposits of $CO_2$ ice, found almost exclusively on their trailing hemispheres. However, best fit spectral models of longer wavelength datasets (~3 – 5 µm) indicate that the spectral signature of $CO_2$ ice is largely absent, and instead, the exposed surfaces of these moons are composed primarily of tiny $H_2O$ ice grains. To investigate possible compositional layering of these moons, we have collected new data using the Infrared Array Camera (IRAC) onboard the Spitzer Space Telescope (~3 – 5 µm). Spectral modeling of these new data is consistent with prior analyses, suggesting that the exposed surfaces of the Uranian moons are primarily composed of tiny $H_2O$ ice grains. Furthermore, analysis of these new data reveal that the trailing hemispheres of these moons are brighter than their leading hemispheres over the 3 to 5 µm wavelength range, except for Miranda, which displays no hemispherical asymmetries in its IRAC albedos. Our analyses also revealed that the surface of Ariel displays five distinct, regional-scale albedo zones, possibly consistent with the spatial distribution of $CO_2$ ice on this moon. We discuss possible processes that could be enhancing the observed leading/trailing albedo asymmetries exhibited by these moons, as well as processes that could be driving the apparent compositional stratification of their near surfaces.

## 1. Introduction

In 1986, Voyager 2 revealed the surfaces of the large and tidally-locked "classical" Uranian moons Miranda, Ariel, Umbriel, Titania, and Oberon for the first time (*e.g.*, Smith et al., 1986, Stone et al., 1986; Croft and Soderblom, 1991) (Table 1). Data returned by the Imaging Science System (ISS, ~0.28 – 0.64 µm) onboard the Voyager 2 spacecraft showed that the surfaces of these moons are darker than the surfaces of $H_2O$ ice-rich Saturnian moons observed previously during the Voyager encounters with Saturn (*e.g.*, Smith et al., 1981, 1982). ISS also showed that the gray-toned surfaces of the Uranian moons have some brighter regions, spatially associated with impact features and tectonized provinces (*e.g.*, Helfenstein et al., 1989; Pappalardo et al., 1997; Beddingfield et al., 2015; Beddingfield and Cartwright, 2019). Subsequent analysis of the ISS data revealed that spectrally red material is present on the classical Uranian satellites, primarily on the *leading* hemispheres (longitudes 1 – 180°) of the outer moons, Titania and Oberon (Buratti and Mosher 1991; Bell et al., 1991; Helfenstein et al., 1991).

Ground-based, near-infrared (NIR) observations (~1 – 2.5 µm) determined that the surface compositions of the Uranian moons are dominated by a mixture of $H_2O$ ice and a dark, spectrally-neutral constituent (*e.g.*, Cruikshank et al., 1977; Cruikshank, 1980; Cruikshank and Brown, 1981; Soifer et al., 1981; Brown and Cruikshank, 1983; Brown and Clark, 1984) that is likely carbonaceous in origin (Clark and Lucey, 1984). The detected $H_2O$ bands on these moons are much weaker than the same $H_2O$ features on ice-rich Saturnian moons (*e.g.*, Cruikshank et al., 1977), demonstrating how the presence of dark material weakens the spectral signature of $H_2O$ ice over these wavelengths (Clark and Lucey, 1984). Overprinting these "dirty" $H_2O$ ice features, narrow $CO_2$ ice bands have been detected (between 1.9 and 2.1 µm), primarily on the *trailing* hemispheres (longitudes 181 – 360º) of the inner moons, Ariel and Umbriel (Grundy et al., 2003, 2006; Cartwright et al., 2015). The central wavelength positions, band shapes, and band strengths of these $CO_2$ ice features are remarkably similar to the second order overtone and combination bands of "pure" $CO_2$ ice (*i.e.*, concentrated deposits of $CO_2$, segregated from other constituents, with crystal structures dominated by $CO_2$ molecules) (*e.g.*, Hansen, 1997; Gerakines et al., 2005).

[1]*Visiting Astronomer at the Infrared Telescope Facility, which is operated by the University of Hawaii under contract NNH14CK55B with the National Aeronautics and Space Administration.*

At longer wavelengths, Miranda and Ariel were observed by the Infrared Interferometer Spectrometer (IRIS) onboard Voyager 2 over the ~20 to 50 µm range (Hanel et al., 1986). Analysis of the IRIS data suggested that these two moons have surfaces composed of isotropically scattering dark grains (Hanel et al., 1986), hinting at their bizarre regolith microstructures. More recently, Ariel, Umbriel, Titania, and Oberon were observed by the Infrared Array Camera (IRAC) onboard the Spitzer Space Telescope (Fazio et al., 2004) in four broad channels spanning ~3.1 to 9.5 µm. Analysis of these IRAC data shows that the Uranian satellites have higher albedos over the 3 to 5 µm range compared to $H_2O$ ice-rich satellites in the Jovian and Saturnian systems (Cartwright et al., 2015). The SpeX spectrograph at NASA's Infrared Telescope Facility (IRTF) (Rayner et al., 1998, 2003), operating in long wavelength cross-dispersed (LXD) mode, was used to collect L/L' band spectra (~2.9 – 4.2 µm) of Ariel, Titania, and Oberon (Cartwright et al., 2018). Analysis of these SpeX/LXD spectra shows that the spectral continua of the Uranian moons are dominated by $H_2O$ ice, with similar continua shapes to the $H_2O$-rich Saturnian moons. However, the Uranian satellites display brighter 3.6-µm peaks compared to the $H_2O$-rich Saturnian moons, and they are also much brighter than dark material-rich moons like Iapetus and Callisto (Fig. 11 in Cartwright et al., 2018). Thus, over the ~0.4 to 2.5 µm wavelength range, the classical Uranian satellites have weaker $H_2O$ ice bands and darker surfaces compared to $H_2O$-rich moons elsewhere, but the Uranian satellites are brighter than other icy moons over long NIR wavelengths (3 – 5 µm).

The mean optical path length (MOPL) estimates the average penetration depth of photons into particulate mixtures (Clark and Roush, 1984). Utilizing the MOPL, we previously calculated the average penetration distance of photons into particulate mixtures of amorphous C and $H_2O$ ice, finding that photons over short NIR wavelengths (1 – 2.5 µm) travel greater distances into these particulate mixtures (~0.15 – > 10 mm depths) compared to photons over long NIR wavelengths (~0.001 – 0.05 mm depths) (Cartwright et al., 2018). Previous radiative transfer modeling work shows distinct differences between these two wavelength regions, with best fit synthetic spectra, spanning the short NIR region, dominated by constituents with ~10 to 50 µm diameter grains, whereas best fit models spanning the long NIR region are primarily composed of grains with ≤ 2 µm diameters (Cartwright et al., 2015, 2018). Other studies have demonstrated that small $H_2O$ ice grains enhance the brightness of the 3.6-µm peak relative to larger grains (*e.g.*, Filacchione et al., 2012), precisely where SpeX/LXD spectra of the Uranian moons show strong continua peaks. Furthermore, short NIR spectral models include areally mixed $CO_2$ ice (10 – 50 µm grain diameters) (Cartwright et al., 2015), but long NIR synthetic spectra include only minor amounts of areally mixed $CO_2$ ice (or none at all), even for Ariel's trailing hemisphere, which is mantled by a large amount of $CO_2$ (Grundy et al., 2003, 2006; Cartwright et al., 2015). Thus, the large grains of $H_2O$ ice, $CO_2$ ice, and other constituents detected in short NIR spectra appear to be masked by a thin veneer of tiny $H_2O$ ice grains in long NIR datasets.

Although previous analyses support this hypothesis, only a small handful of IRAC and SpeX/LXD observations of the Uranian satellites have been published. Consequently, the spectral properties of these moons at wavelengths > 2.5 µm are still poorly constrained and follow-up analyses are needed to explore whether their regoliths are compositionally stratified. In this study, we investigate the spectral properties of these moons over the long NIR region, using new IRAC channel 1 (Ch.1) and channel 2 (Ch.2) observations of Ariel, Umbriel, Titania, and Oberon. We also report IRAC Ch.1 observations of Miranda, which represent the first published data for this moon over the 3 to 5 µm wavelength range. We utilize a Hapke-Mie hybrid radiative transfer model to guide our interpretation of the constituents mantling the surfaces of these moons.

## 2. Spitzer/IRAC Observations and Data Reduction

IRAC has been gathering NIR spectrophotometry of astronomical targets since its launch in 2003 (Werner et al., 2004). During the cryogenically-cooled phase of Spitzer's lifespan, IRAC collected images in four broadband channels, centered near 3.6, 4.5, 5.8, and 8.0 µm (Fazio et al., 2004). The detector arrays for each channel are 256 x 256 pixels, with a pixel size of 1.2" x 1.2". The field of view (FOV) of each channel is 5.8 x 5.8 arcmin, with images collected in pairs that share the same FOV (channels 1 and 3, and channels 2 and 4, respectively). As one pair of channels observes the target, the other two channels observe off-target sky, with no overlap in FOV. Previous IRAC observations of the Uranian moons were conducted using all four channels (Program 71, 2003 – 2005; Cartwright et al., 2015). Since 2009, Spitzer has transitioned to the passively-cooled phase of its lifespan, and Ch.1 and Ch.2 (centered near 3.6 and 4.5 µm) are still operable at comparable sensitivities to Spitzer's cryogen-cooled phase.

In 2015, we collected new IRAC Ch.1 and Ch.2 images of the Uranian satellites (Program 11112). These data were collected as sets of nine dithered images in each channel (26.8 s exposures). IRAC observed the leading and trailing hemispheres of these moons two times in each channel (primary and follow-up observations). During the targeted observations of each moon's leading and trailing hemisphere, the other moons were also visible in the FOV. Consequently, IRAC effectively imaged each moon sixteen times in both channels (observations summary shown in Table 2).

We analyze IRAC corrected basic calibrated data (CBCD) products in this study, which are dark-subtracted, flatfielded, and flux-calibrated using the standard Spitzer Science Center (SSC) data reduction pipeline (see IRAC handbook[2] for more information). The SSC pipeline also removes common IRAC artifacts, including: mux-stripe, column pulldown, banding, saturation, and stray light contributions. Using corrections supplied by SSC, we corrected for variations in the pixel solid angle and for photometric variations across the array in all analyzed data. After performing these corrections, all analyzed images were converted into units of mJy/pixel.

[2]http://irsa.ipac.caltech.edu/data/SPITZER/docs/irac/iracinstrumenthandbook/

Next, we performed aperture photometry, using a 3-pixel radius aperture centered on each moon for all nine dithered frames. An annulus of pixels surrounding the aperture was utilized for background subtraction (inner and outer radii of 3 and 7 pixels). A 3-pixel aperture allows us to maximize flux extraction for each moon while minimizing sky background contributions, cosmic ray hits, and other unwanted flux contributions. To correct for the difference in radius between our 3-pixel apertures and the SSC-calibrated 10-pixel aperture, we multiplied our flux estimates by channel-dependent aperture corrections. To account for the solar spectral slope, we divided our flux estimates by solar color corrections (from Smith and Gottlieb, 1974). Absolute flux calibration for IRAC is accurate to 2% (Reach et al., 2005). Flux uncertainties include differences between each of the nine CBCD frames in each observation, the chosen aperture and annulus size we used for flux extraction, and photon counting statistics. We averaged the extracted fluxes for all nine frames and converted these mean fluxes into geometric albedos using the same routines reported in Section *2.2* of Cartwright et al. (2015). These routines utilize the observation viewing geometry, target radius, target heliocentric distance, observer-target distance, and additional Uranian satellite-specific photometric parameters originally reported in Karkoschka (2001).

Some of the IRAC observations reported here were contaminated by scattered light from Uranus or from a neighboring moon. To correct for these sources of scattered light, we utilized different background subtraction routines. For satellite observations contaminated by scattered light from Uranus, we subtracted off an annulus of pixels that contain only background flux and scattered light, centered at the same distance from Uranus as the contaminated moon. To correct frames where two proximal moons have convolved fluxes, we generated median flux models for each moon, and then subtracted these models from each contaminated frame, thereby separating their fluxes. We tested these scattered light removal routines on non-contaminated observations, finding good agreement with the results of our standard background subtraction routines. Additionally, we applied these scattered light removal routines to five previously unreported Program 71 observations of these moons (Table 2). After removing scattered light from all contaminated frames, we averaged the corrected fluxes and converted them into geometric albedos (Section *2.2* of Cartwright et al., 2015).

## 3. Results and Analysis

### *3.1 IRAC Photometry*

We report 75 Ch.1 and 65 Ch.2 fluxes and geometric albedos for the Program 11112 observations, and five new Ch.1 and Ch.2 measurements for the Program 71 observations of the Uranian satellites (Table 3, Figure 1). To facilitate comparison with prior IRAC results, we show the previously published Program 71 measurements in Table 3 as well. These new results include six Ch.1 observations of Miranda. The six corresponding Ch.2 observations of Miranda, as well as four other Ch.2 observations of Ariel, are heavily contaminated by scattered light from Uranus, even after application of our scattered light removal routines. Consequently, we do not report those ten Ch.2 fluxes and albedos for Miranda and Ariel here, and we exclude them from subsequent analysis.

Amongst the five classical moons, Ariel has the highest albedo and displays the greatest amount of variation across its surface, with higher albedos measured over its trailing hemisphere (Ch.1, 0.203 to 0.236 ± *0.002 to 0.007*) compared to its leading hemisphere (Ch.1, 0.174 to 0.203 ± *0.002 to 0.006*). Miranda is the next brightest moon (Ch.1, 0.190 to 0.202 ± *0.016 to 0.024*), followed by Oberon (Ch.1, 0.163 to 0.169 ± *0.001 to 0.002*), Titania (Ch.1, 0.157 to 0.163 ± *0.001 to 0.002*), and Umbriel (Ch.1, 0.146 to 0.163 ± *0.003 to 0.008*). These four moons display more subtle variations between their leading and trailing hemispheres compared to Ariel. The relative brightness of these moons is broadly consistent with their shorter wavelength albedos (Table 1). However, at shorter wavelengths, Titania is brighter than Oberon (0.39 ± *0.02* and 0.33 ± *0.01*, respectively, Table 1), whereas over IRAC wavelengths, Oberon is brighter than Titania.

### *3.2 Mean IRAC Albedos*

#### *3.2.1 Leading vs. Trailing Hemispheres*

To determine whether these moons display longitudinal asymmetries in brightness, we calculated mean Ch.1 and Ch.2 geometric albedos for each moon's leading and trailing hemisphere (Table 4, Figure 2) and propagated errors (Appendix A). Previously reported mean IRAC albedos (Cartwright et al., 2015) and SpeX/LXD albedos (Cartwright et al., 2018) are included in Table 4 to facilitate comparison between these different long NIR datasets.

Our results show clear differences (> 3$\sigma$) between the albedos of Ariel's leading and trailing hemispheres, as well as smaller, but apparent, differences (> 2$\sigma$) between the leading and trailing hemispheres of Titania and Oberon. Umbriel displays more subtle differences between the Ch.1 (< 2$\sigma$) and Ch.2 (< 1$\sigma$) albedos of its leading and trailing sides. We detect no discernable differences between Miranda's leading and trailing hemispheres. Intercomparison of these moons' albedos demonstrates that Ariel's trailing hemisphere is substantially brighter than any region on Umbriel, Titania, or Oberon (> 3$\sigma$), and is possibly brighter than any region on Miranda (> 1$\sigma$). Thus, the trailing hemispheres of Ariel, Titania, and Oberon (and probably Umbriel) are brighter than their leading hemispheres over long NIR wavelengths. We discuss the possible processes that could be enhancing the albedos of these moons' trailing hemispheres in section *4.1*.

#### *3.2.2 Southern vs. Northern Hemispheres*

To determine whether these moons display latitudinal variations in brightness, we compared the Program 11112 IRAC albedos (subsolar latitudes ~32.2 – 33.9ºN) to the Program 71 albedos (subsolar latitudes ~6.7 – 18.3ºS) (Table 4, Figure 3). In general, we find that the measured albedos for the leading and trailing hemispheres of these moons are similar across their southern and northern latitudes. A possible exception is Ariel's trailing hemisphere, which appears to be brighter over its southern latitudes compared to its northern latitudes (> 1$\sigma$ difference). However, this difference is likely spurious, as discussed in the following subsection.

#### *3.2.3 Identification of Five Distinct Albedo Zones on Ariel*

Visual inspection of our results (Figure 1) suggests that additional, regional-scale albedo variations are present on Ariel. To investigate this possibility further, we separated Ariel's individual albedo measurements into different longitudinal zones, averaged them together, and propagated errors (Appendix A1). The five resulting albedo zones are defined as follows: "Uranus-facing" (UF),

“leading” (LE), “anti-Uranus” (AU), “peripheral-trailing” (PT), and “central-trailing” (CT). The number of data points and range of longitudes in each of these albedo zones are summarized in Table 6. Using the same range of longitudes, we reclassified the previously reported Program 71 albedos for Ariel, finding that these older data were collected entirely over the LE and CT albedo zones (Table 5). The Ch.1 albedos of each of the five albedo zones display > $2\sigma$ differences compared to their adjacent zones, except for the AU zone (> $1\sigma$ difference).

Comparison of these albedos zones (for both Ch.1 and Ch.2) shows that Ariel is brightest near the center of its trailing hemisphere (CT zone), less bright outside of this central region (PT zone), darker over transitional longitudes (AU and UF zones), and darkest near the center of its leading hemisphere (LE zone). Furthermore, Ariel’s CT zone displays enhanced albedos over both southern (Program 71) and northern (Program 11112) latitudes, demonstrating the large spatial extent of this bright region. The spatial trends in brightness across Ariel’s surface suggests that its trailing hemisphere is mantled by a reflective material, peaking in abundance near its antapex (longitude 270º). This trend in brightness is likely consistent with the spatial distribution of $CO_2$ ice (Grundy et al., 2003, 2006; Cartwright et al., 2015), suggesting that the presence of $CO_2$ is enhancing the IRAC albedos of Ariel’s trailing side. Supporting this interpretation, spectroscopic laboratory measurements demonstrate that $CO_2$ ice (*e.g.*, Hansen et al., 1997) is much brighter than $H_2O$ ice (*e.g.*, Mastrapa et al., 2009) over the ~3 to 4 µm and ~4.5 to 5 µm wavelength ranges covered by IRAC.

Identification of these five albedo zones on Ariel demonstrates that the previously described differences between the Program 71 and 11112 observations of Ariel’s trailing hemisphere (section *3.2.2*) are no longer valid. IRAC observed both the PT and CT albedo zones during Program 11112, but only the brighter CT zone during Program 71. Consequently, the lower mean albedo calculated using the Program 11112 observations results from the averaging of two distinct albedo zones on Ariel’s trailing side. When we compare only the CT zone measurements, the mean albedos for the Program 11112 and 71 datasets are essentially identical. However, over the center of Ariel’s leading hemisphere (LE zone), there is a > $1\sigma$ difference between these two datasets, with slightly brighter results for the Program 71 observations. Analysis of short NIR spectra suggest that more $H_2O$ ice is exposed over Ariel’s northern latitudes compared to its southern latitudes (Cartwright et al., 2018). Consequently, perhaps there is more exposed $H_2O$ ice over the northern latitudes of Ariel’s leading hemisphere compared to the southern latitudes of its leading side, resulting in greater absorption and lower IRAC albedos. This exposed $H_2O$ ice could be concentrated at Ariel’s north pole, which may be denuded of $CO_2$ ice and other volatiles, as predicted by thermodynamical models (Grundy et al., 2006; Sori et al., 2017).

After comparing the Program 71 and 11112 results, we combined these two datasets into one set of albedo zones (bolded columns in Table 5, Figure 4). Additionally, we compared these IRAC results to the two published SpeX/LXD albedos of this moon (Cartwright et al., 2018), which were collected over Ariel’s UF and CT albedo zones. This comparison demonstrates broad agreement between these SpeX/LXD and IRAC albedos (Table 5).

### *3.3 Model Fitting IRAC Light Curves*

As described in section *3.2.1*, the trailing hemispheres of the Uranian moons are brighter than their leading hemispheres. To further investigate these longitudinal trends, we fit the IRAC albedos for each moon, weighted by their uncertainties, with mean and sinusoidal models (Figure 5). The mean models represent surfaces with no discernable longitudinal variations in albedo, whereas the sinusoidal models represent surfaces with significant longitudinal variations in albedo. After fitting the data with these two different models, we compared the fits using an $F$-test (*e.g.*, Speigel, 1992). The null hypothesis for this $F$-test is that there is no difference between the fits provided by the mean and sinusoidal models. Because only six IRAC observations exist for Miranda, which sample only a small range of longitudes on its leading (~78 – 84º) and trailing (~254 – 259º) hemispheres (Figure 1a), we exclude Miranda from this analysis.

For Ariel, Umbriel, Titania, and Oberon, the $F$-test results demonstrate that the sinusoidal model provides a significantly better fit for the Ch.1 ($p < 0.009$) and Ch.2 ($p < 0.0003$) light curves compared to the mean model (Table 6). These results indicate that the trailing hemispheres of these moons are brighter than their leading hemispheres, consistent with our mean IRAC albedo analyses (section *3.2.1*). Of note, this approach only considers whether a mean or sinusoidal model is a better fit to the measured albedos, and it does not allow for formal exclusion of either model as a reasonable fit to the data.

## 3.4 Radiative Transfer Modeling

### *3.4.1 Best Fit Synthetic Spectra*

We conducted radiative transfer modeling to explore the composition and microstructure of the Uranian satellites' regoliths. These spectral models utilize a hybrid Hapke-Mie approach (Appendix A2), which has been applied previously to IRAC and SpeX datasets of these moons (Cartwright et al, 2015, 2018). We generated best fit synthetic spectra for each moon's leading and trailing hemisphere and for Ariel's five albedo zones (Tables 7 and 8, Figure 6), making sure that all the synthetic spectra reproduce the measured Ch.1 and Ch.2 albedos. These best fit models are composed of intimate mixtures of $H_2O$ ice (multiple grain sizes) and dark material, with no $CO_2$ ice included.

We utilized the previously collected SpeX/LXD spectra of Ariel, Titania, and Oberon to provide an additional control on the shape of the continua for the spectral models of these three moons (Figure 6). We calculated reduced Chi Square ($\chi^2$) statistics (*e.g.*, Bevington and Robinson, 1969) to assess the goodness-of-fit between the synthetic spectra and the SpeX/LXD spectra. In general, a reduced $\chi^2$ value > 1 indicates a poor fit between a model and the observed data, whereas a reduced $\chi^2$ value < 1 indicates that a model is likely a good fit to the observed data. Model fitting of datasets with large uncertainties can lead to erroneously low $\chi^2$ values. Given the low signal-to-noise (S/N) of the SpeX/LXD spectra, the $\chi^2$ values reported here could be underestimated. Consequently, both visual assessment and reduced $\chi^2$ values were used to assess the quality of fits between the observed data and the spectral models.

We compared these best fit synthetic spectra to identify possible compositional trends on these satellites. The spectral models for all five moons include substantial fractions (> 50%) of tiny $H_2O$ ice grains (1 – 2 µm diameters), as well as a small amount (≲ 1%) of sub-micron diameter $H_2O$ ice grains. Additionally, all of the synthetic spectra include a modest fraction (~4 – 28%) of larger $H_2O$ ice grains (10 µm diameters), except for the models of Ariel, which are dominated by sub-micron to 2 µm diameter $H_2O$ ice grains (~97 – 99%). All of these models include some amount of low albedo constituents (amorphous C and Triton tholins, ~1 – 10 µm diameters), with less dark material included in the models of the brighter inner moons Ariel and Miranda (~1 to 3%), and substantially more dark material included in the models of the outer moons Umbriel, Titania, and Oberon (~13 to 33%). Therefore, the synthetic spectra presented here suggest that the Uranian satellites' regoliths are primarily composed of tiny $H_2O$ ice grains (~1 – 2 µm diameters). The prevalence of 1 to 2 µm diameter $H_2O$ grains is consistent with previous spectral modeling efforts (Cartwright et al., 2015, 2018). We explore how the surfaces of these moons might have developed thin mantles of tiny $H_2O$ ice grains in section *4.2*.

*3.4.2 Spectral Modeling of Ariel: Investigating the Influence of $CO_2$ Ice*

Although our best fit synthetic spectra contain no $CO_2$ ice, spectral observations made by SpeX in short cross-dispersed (SXD) mode clearly demonstrate that $CO_2$ ice is present on the trailing hemispheres of these moons, in particular on Ariel (Grundy et al., 2003, 2006; Cartwright et al., 2015). Given the clear evidence for $CO_2$ ice at shorter wavelengths (27% of best fit spectral models), perhaps the high IRAC albedos for Ariel's trailing hemisphere are influenced by the presence of $CO_2$ ice. To investigate this possibility further, we generated a suite of spectral models that include $CO_2$ ice. We focused this modeling effort on Ariel's bright central-trailing (CT) zone. Examples of the $CO_2$-included models we generated are reported in Table 9 and shown in Figure 7. These synthetic spectra include examples of areally mixed $CO_2$ ice (Figure 7a) and intimately mixed $CO_2$ ice (Figure 7b). For reference, we include the best fit model for the SpeX/SXD spectra of Ariel's trailing hemisphere, extrapolated over IRAC wavelengths (Figure 7a).

The SpeX/SXD best fit model cannot reproduce the IRAC albedos and has a reduced $\chi^2$ value > 1. Visual assessment of this model demonstrates that it provides a poor fit to the SpeX/LXD spectrum. The example spectral models that include areally mixed $CO_2$ ice have reduced $\chi^2$ values < 1, suggesting reasonable fits, but visual assessment shows that they provide poor fits to the SpeX/LXD spectrum between 3 and 3.4 µm. Additionally, these spectral models include less areally mixed $CO_2$ ice (13 – 14%) compared to the SpeX/SXD best fit model (27%). The two example spectral models that include intimately mixed $CO_2$ also have reduced $\chi^2$ values < 1 and provide much better fits to the SpeX/LXD spectral continuum between 3 and 3.4 µm. Furthermore, the two intimately mixed $CO_2$-included models include a sizable faction of $CO_2$ ice (19 and 27%). Thus, our $CO_2$-included spectral modeling efforts demonstrate that IRAC could be sensing modest abundances of $CO_2$ ice, but that this constituent is most likely intimately mixed with $H_2O$ ice and dark material. In contrast, SpeX/SXD best fit models clearly indicate the presence of areally mixed $CO_2$ ice. We explore the possible role of tiny $H_2O$ ice grains in obscuring areally mixed $CO_2$ ice in section *4.2.4*.

*3.4.3 Spectral Modeling of Ariel: Investigating the Influence of $H_2O$ Ice*

To investigate the influence of $H_2O$ ice grain size on these moons, we generated a suite of pure $H_2O$ ice spectral models (1, 10, and 100 µm grain diameters) (Table 10). Because Ariel's trailing hemisphere is brighter than the other moons, we focused this analysis on Ariel's central-trailing (CT) zone. For reference, we compare the best fit spectral model of Ariel's CT zone, which includes no $CO_2$ ice (described in Table 8), to these pure $H_2O$ ice synthetic spectra (Figure 7c).

Visual inspection demonstrates that the 100 µm $H_2O$ ice synthetic spectrum provides poor fits to the IRAC albedos and to the shape of the SpeX/LXD continuum, and this model has a reduced $\chi^2$ value > 1. The 10 µm $H_2O$ model provides a reasonable match to the IRAC Ch.1 albedo, but not the Ch.2 albedo, and it provides a poor fit to the SpeX/LXD continuum (albeit, with a $\chi^2$ statistic < 1). The 1 µm $H_2O$ model provides a much better fit to the SpeX/LXD continuum compared to the 10 and 100 µm $H_2O$ models, it has a $\chi^2$ statistic < 1, and provides a reasonable match to the Ch.1 albedo, but not the Ch.2 albedo, for Ariel's CT zone. Comparison of these pure $H_2O$ models demonstrates that tiny $H_2O$ ice grains can match the shape of the 3.6-µm peak but larger $H_2O$ grains cannot, highlighting the influence of tiny grains on the spectral signature of these moons.

## 4. Discussion

### *4.1 Leading/Trailing Asymmetries in IRAC Albedos*

#### *4.1.1 Driven by Heliocentric Dust Impacts?*

Assuming the Uranian moons are $H_2O$ ice-rich over the range of depths probed by dust impacts (top few cm), the lower IRAC albedos and stronger $H_2O$ ice bands on their leading hemispheres could result from enhanced regolith overturn driven by heliocentric micrometeorite impacts. This process should expose "fresher," less space-weathered $H_2O$ ice on icy satellites, and perhaps bury previously exposed tiny $H_2O$ ice grains as well, thereby enhancing $H_2O$ ice band strengths (*e.g.*, Bennett et al., 2013). Because of gravitational focusing by Uranus, heliocentric dust particles, entering the Uranian system at high velocities (~30 km/s), will collide more frequently with the inner moons compared to the outer moons (Tamayo et al., 2013). Additionally, the faster orbital velocities of the inner moons Miranda and Ariel (6.7 and 5.5 km/s, respectively) should increase the frequency of collisions between heliocentric dust particles and their leading hemispheres compared to the outer moons Umbriel, Titania, and Oberon (4.5, 3.6, and 3.2 km/s, respectively). The large leading/trailing asymmetry in $H_2O$ ice band strengths and IRAC albedos on the inner moon Ariel, and the subtle hemispherical asymmetries observed on the more distant moons Umbriel, Titania, and Oberon, appears to be consistent with enhanced regolith overturn of their leading hemispheres by heliocentric dust impacts (Cartwright et al., 2018).

If heliocentric dust collisions are driving the leading/trailing asymmetries in these measurements, then the innermost moon Miranda should exhibit similar hemispherical trends due to its high orbital velocity and proximity to Uranus. However, the results presented here, along with previous analyses (Cartwright et al., 2018), indicate that Miranda does *not* display obvious leading/trailing asymmetries in either its IRAC albedos or $H_2O$ ice band strengths. Additionally, although dust collisions could promote burial of existing tiny $H_2O$ grains, they could also promote grain fragmentation via impact comminution, thereby generating new tiny grains of $H_2O$ ice. Thus, comparison of the hemispherical trends in composition on Ariel, Umbriel, Titania, and Oberon to the absence of hemispherical trends on the innermost moon Miranda suggests that heliocentric dust impacts are not the primary driver of stronger $H_2O$ ice bands and lower IRAC albedos on the leading hemispheres of these moons.

Alternatively, perhaps $H_2O$ ice bands on Miranda's trailing hemisphere are being enhanced by another process, complementing enhancement of $H_2O$ ice bands on its leading hemisphere by micrometeorite impacts. It has been previously suggested that ring particles could become electrically charged and experience a Lorentz force from Uranus' rotating magnetic field (Grundy et al., 2006). Such a process would tend to push larger grains inward on decaying orbits and tiny grains outward on expanding orbits. Uranus' outermost µ-ring is dusty and diffuse, with particle orbits stretching from ~86,000 to 103,000 km (Showalter and Lissauer, 2006). Peak particle densities in the µ-ring correspond to the orbit of the ring moon Mab, which likely represents the primary source of µ-ring particles (*e.g.*, Showalter and Lissauer, 2006; de Pater et al., 2006). The µ-ring has a strong blue color, consistent with a population of tiny $H_2O$ ice grains that have sub-micron to micron diameters (de Pater et al., 2006).

Tiny $H_2O$ grains in the µ-ring could spiral outward on expanding orbits after becoming electrically charged by interactions with charged particles in Uranus' magnetosphere. The orbits of these electrically charged $H_2O$ grains would eventually cross the orbit of the innermost classical moon Miranda ($a$ ~130,000 km) and subsequently collide with its trailing hemisphere. The impact velocities of these collisions are likely low, and it is unclear whether they would tend to promote regolith overturn or preferentially mantle Miranda's trailing hemisphere. Nevertheless, collisions between µ-ring particles and Miranda's trailing hemisphere, and collisions between heliocentric dust impacts and Miranda's leading hemisphere, could be promoting global-scale regolith overturn, thereby erasing hemispherical asymmetries in its composition.

#### *4.1.2 Driven by the Presence of $CO_2$ Ice?*

It has been hypothesized that the $CO_2$ ice detected on these moons is part of a radiolytic production cycle, whereby $CO_2$ molecules are generated by magnetospherically-embedded charged particle irradiation of native $H_2O$ ice and C-rich species (Grundy et al., 2006; Cartwright et al., 2015). Radiolytic generation of $CO_2$ molecules has been observed in numerous laboratory studies, utilizing different irradiation sources (heavy ions, protons, electrons, and UV photons), different substrates made of $H_2O$ ice and various C-rich materials, and over a wide range of cryogenic temperatures (~10 – 150 K) relevant to icy objects in the outer Solar System (*e.g.*, Strazzulla and Palumbo, 1998; Chakarov et al., 2001; Gerakines and Moore, 2001; Hudson and Moore, 2001; Mennella et al., 2004; Sedlacko et al., 2005; Gomis and Strazzulla, 2005; Loeffler et al., 2005; Jamieson et al., 2006; Mennella et al., 2006; Kim and Kaiser, 2012; Raut et al., 2012).

$CO_2$ molecules generated *in situ* by charged particle irradiation on the Uranian satellites should sublimate from radiolytic production sites, hop along the surfaces of these moons, and then merge with deposits of concentrated $CO_2$ that likely build up on colder, reflective regions, such as bright, $H_2O$-rich crater floors and tectonic landforms like chasmata (Grundy et al., 2006; Sori et al., 2017). These $CO_2$-rich deposits are likely thermodynamically stable over long timescales (~0.01 – 1 Ga; Grundy et al., 2006; Sori et al., 2017). Consequently, $CO_2$-rich deposits likely expand and become thicker over time as they accumulate more migrating $CO_2$ molecules.

Although initially made up of small $CO_2$ grains, these deposits could eventually sinter into slabs of $CO_2$ ice. Analysis of $CO_2$ ice in the laboratory suggests that the second order combination and overtone bands (between 1.9 and 2.1 µm) of this constituent require relatively thick deposits in order to detect and characterize them (~1 – 10 mm thick samples) (*e.g.*, Hansen, 1997, 2005; Gerakines et al., 2005). Third order $CO_2$ bands (between 1.57 and 1.61 µm) are a factor of ~60 to 200 weaker than second order $CO_2$ bands, requiring samples up to ~100 mm thick to characterize in the laboratory (Hansen, 2005). Second order $CO_2$ bands have been detected in spectra of the Uranian moons, and third order $CO_2$ bands have also been detected in spectra of Ariel (Grundy et al., 2006). Consequently, the weak $CO_2$ ice bands detected on these moons support the presence of thick deposits of $CO_2$ ice. These thick $CO_2$ ice deposits could weaken the spectral signature of the underlying $H_2O$ ice-rich cold traps, thereby weakening $H_2O$ ice bands and increasing the IRAC albedos of these moons' trailing hemispheres compared to their leading hemispheres, where $CO_2$ is largely absent.

The large reduction in $H_2O$ ice band strengths and increase in IRAC albedos on Ariel's trailing hemisphere is consistent with the large abundance of $CO_2$ ice on its trailing hemisphere (27% of best fit spectral models; Cartwright et all., 2015). The similar, but more subtle, spatial trends in $H_2O$ ice band strengths and IRAC albedos on the more distant moons Umbriel, Titania, and Oberon is also consistent with the distribution of $CO_2$ ice on their trailing hemispheres (8%, 5%, and 3% of best fit spectral models, respectively; Cartwright et al., 2015). Miranda displays no hemispherical asymmetries in its $H_2O$ ice band strengths or IRAC albedos, and $CO_2$ ice has *not* been detected on this moon (Bauer et al., 2002; Grundy et al., 2006; Gourgeot et al., 2013; Cartwright et al., 2015, 2018). Miranda's low mass (Table 1) likely leads to efficient Jean's escape of radiolytically-generated $CO_2$ molecules, effectively depleting Miranda's surface inventory of $CO_2$ (Sori et al., 2017). Therefore, the presence of $CO_2$ ice could explain the higher IRAC albedos and weaker $H_2O$ ice bands on the trailing hemispheres of Ariel, Umbriel, Titania, and Oberon, and the absence of $CO_2$ on Miranda could explain the lack of hemispherical trends in its composition.

Although the spectral signature of areally mixed $CO_2$ ice is apparent in short NIR spectra, it is absent from long NIR datasets, possibly obscured by tiny $H_2O$ ice grains. Spectral modeling of Ariel's central-trailing (CT) albedo zone indicates that IRAC could be sensing large fractions of intimately mixed $CO_2$ ice, along with tiny $H_2O$ grains (section *3.4.2*). Perhaps intimately mixed $CO_2$ ice is increasing the IRAC albedos of Umbriel's, Titania's, and Oberon's trailing hemispheres as well. We consider a possible mechanism for obscuring the spectral signature of areally mixed $CO_2$ ice deposits on these satellites' surfaces in section *4.2.4*.

### *4.2 Probing the Microstructure of the Uranian Satellites' Regoliths*

The apparent disconnect between the spectral signatures of the Uranian satellites in SpeX/SXD (~1 – 2.5 µm) and Spitzer/IRAC and SpeX/LXD (~3 – 5 µm) datasets has been interpreted to result from differences in photon penetration depths into $H_2O$ ice-rich substrates over these two wavelength regions (Cartwright et al., 2015, 2018). Therefore, these datasets could be sampling different compositional layers, with the IRAC and SpeX/LXD data primarily sampling shallower depths (top ~0.001 to 0.05 mm), whereas the SpeX/SXD spectra are able to probe deeper depths (~0.15 – 10 mm) into the regoliths of these moons. Supporting this interpretation, ground-based polarimetric observations, made over visible (VIS) wavelengths, indicate that the classical Uranian moons display strong negative polarization branches (NPBs) at low phase angles (Afanasiev et al., 2014). These NPBs are much steeper than both bright, $H_2O$ ice-rich moons, like Tethys and Rhea, and dark material-rich moons, like Callisto and Iapetus, measured over comparable phase angles (*e.g.*, Kiselev et al., 2017). The distinct VIS polarization properties of the Uranian satellites are consistent with the presence of tiny grains ($\lesssim$ 1 μm diameters) forming a porous and "crumbly" layer (Afanasiev et al., 2014). Recent laboratory experiments that investigated the polarization properties of $H_2O$ ice support this interpretation, showing that small $H_2O$ ice grains generate more pronounced NPBs than large $H_2O$ grains (Poch et al., 2018).

Our IRAC and SpeX/LXD spectral models, as well as analysis of VIS polarimetric results, suggest that a veneer of tiny grains mantle the surfaces of the Uranian satellites. We cannot, however, discern whether this layer is exclusively made up of tiny grains, or whether it includes larger grains as well ($\gtrsim$ 10 μm diameters). It is also possible that tiny grains are present beneath this exposed layer, filling in pore spaces between larger grains. Consequently, the precise physical structure of these moons' regoliths remains uncertain. In the following subsections, we discuss two different mechanisms that might generate porous layers with large abundances of tiny grains. We also consider possible explanations for why similar veneers of tiny $H_2O$ grains have not been detected on Jovian and Saturnian icy moons.

#### *4.2.1 Regolith Microstructure Developed by Micrometeorite Impacts?*

A porous layer of tiny grains could have been formed via dust collisions with the Uranian moons' surfaces. In this scenario, tiny grains of $H_2O$ ice and other constituents were delivered to the Uranian satellites by different sources of micrometeorites, including heliocentric dust, μ-ring particles (primarily delivered to Miranda), and dust liberated from the surfaces of the Uranian irregular satellites (primarily delivered to Titania and Oberon) (*e.g.*, Tamayo et al., 2013). Although the surfaces of Miranda and Ariel have heavily tectonized regions that are relatively young (~0.1 – 1.0 Ga), they also display heavily cratered, ancient regions (~2 – 3.5 Ga) (*e.g.*, Zahnle et al., 2003). The three other moons, Umbriel, Titania, and Oberon, have heavily cratered surfaces that are all very ancient (~4 – 4.5 Ga), with some younger regions observed on Titania (~2 Ga; Zahnle et al., 2003). The ancient surfaces, or at least the older regions, of these moons could have developed mantles of tiny heliocentric and planetocentric dust grains that have built up over the past few Gyr. Additionally, these different sources of dust could have fragmented exposed grains of $H_2O$ ice as they collided with these moons, thereby increasing the number of tiny $H_2O$ grains on their surfaces. These tiny grains are able to persist on the Uranian moons due to their low surface temperatures (~30 – 90 K; Sori et al., 2017), which reduce the rate of $H_2O$ grain growth via sintering.

#### *4.2.2 Regolith Microstructure Resulting from Volatile Loss?*

Another possible formation mechanism for a porous layer dominated by tiny $H_2O$ grains could involve outgassing of volatiles that originally accreted into these satellites as they formed in the Uranian subnebula (*e.g.*, Lewis, 1972, 1973; Prinn and Fegley, 1981,1989). In this scenario, substantial amounts of intimately mixed C-rich volatiles like $CH_4$ clathrates ($CH_4 \cdot 6H_2O$) were efficiently removed by sublimation and/or radiolytically modified into refractory C-rich residues.

The resulting regolith, composed of $H_2O$ ice and C-rich residues, formed a porous upper layer, which was further comminuted by dust collisions. The ancient surfaces of the Uranian satellites allowed this volatile-depleted layer to persist, at least in some regions, over the age of these moons.

Mid-sized trans-Neptunian objects (TNOs) likely experienced substantial volatile loss via outgassing and subsequent Jeans escape of volatile molecules (Schaller and Brown, 2007; Brown, 2012). In contrast, more massive TNOs like Eris, Pluto, and the large Neptunian moon Triton have retained a much larger fraction of their original volatile inventory. Similar to the classical Uranian satellites, mid-sized TNOs like Orcus, Ixion, Huya, Varuna, and 1999 DE9 have dark surfaces that are possibly composed of carbonaceous residues mixed with different amounts of $H_2O$ ice (*e.g.*, de Bergh et al., 2013, and references therein). As observed during the New Horizons flyby of Pluto, Charon is also primarily composed of $H_2O$ ice mixed with C-rich material (e.g., Grundy et al., 2016). These mid-sized, $H_2O$ ice-bearing TNOs have estimated diameters ranging from ~400 to 1200 km, which are comparable to the diameters of the Uranian satellites (~500 – 1500 km). Furthermore, the polarization properties of these mid-sized TNOs are similar to the Uranian satellites, with steep NPBs that are clearly distinct from large and bright TNOs, which have fairly constant NPBs (*e.g.*, Bagnulo et al, 2008, 2011; Belskaya et al., 2012; Afanasiev et al., 2014). The similarly steep NPBs of these TNOs and the Uranian moons could stem from enhanced scattering in porous layers composed primarily of tiny $H_2O$ grains, formed (at least in part) by volatile outgassing. Volatile outgassing could represent an important process for shaping the regoliths of small and mid-sized icy objects throughout the outer Solar System.

*4.2.3 Comparison to the Jovian and Saturnian Satellites*

Analysis of long NIR spectra indicates that $H_2O$ ice-rich Saturnian moons, and dark material-rich Iapetus and Callisto, are not mantled by porous layers composed primarily of tiny $H_2O$ grains, at least not to the same extent as the Uranian moons. The surfaces of most of the Jovian and Saturnian icy moons are thought to be younger than the Uranian satellites (*e.g.*, Zahnle et al., 2003), reducing the amount of time for layers of tiny grains to develop due to dust impact comminution. Additionally, Jovian and Saturnian satellites have warmer surfaces (~100 – 160 K) than the Uranian moons (~30 – 90 K; Sori et al., 2017), which should promote more rapid $H_2O$ ice grain sintering and the removal of tiny grains. Even the ancient surfaces of Callisto and Iapetus are unlikely to build up layers of tiny $H_2O$ grains due to their high peak temperatures (~130 – 160 K) spurring $H_2O$ ice sintering (*e.g.*, Boxe et al., 2007). Furthermore, the Jovian and Saturnian magnetospheres are more intense than the Uranian magnetosphere, with larger populations of high-energy charged particles (*e.g.*, Cassidy et al., 2010, and references therein). Bombardment by magnetospherically-embedded particles, in particular heavy ions, will tend to preferentially remove small grains via sputtering (*e.g.*, Johnson et al., 2013, and references therein). Measurements made during the Voyager 2 flyby indicate that heavy ions are largely absent from the Uranian magnetosphere (*e.g.*, Ness et al., 1986; Stone et al., 1986). If the absence of heavy ions has persisted over geologic timescales, then perhaps charged particle sputtering of tiny grains is relatively inefficient in the Uranian system. Therefore, the mostly younger surfaces, warmer temperatures, and more frequent magnetospheric interactions could efficiently remove small grains from the surfaces of the Jovian and Saturnian satellites, preventing them from building up veneers of tiny $H_2O$ ice grains.

*4.2.4 Obscuring the Spectral Signature of $CO_2$ Ice*

$CO_2$ molecules on the Uranian moons are likely formed via a radiolytic production cycle, migrating to concentrated deposits of $CO_2$ that sinter into thick $CO_2$ ice slabs. This process presumably occurs more rapidly during winter, when the poles of these moons remain in non-interrupted darkness for over 21 years and temperatures plummet to an estimated 30 K (Sori et al., 2017). Once winter ends, $CO_2$ ice deposits are again exposed to sunlight and surface temperatures increase, perhaps driving a solid-state greenhouse effect (*e.g.*, Matson and Brown, 1989).

In this scenario, $CO_2$ ice deposits are exposed to optical wavelength photons, which penetrate through the translucent, non-scattering $CO_2$ slabs. These photons are absorbed by $H_2O$ ice and dark material at the interface with the bottom of the slabs and are reemitted at thermal wavelengths. These longer wavelength photons are unable to efficiently escape back through the $CO_2$ ice layer, and instead are absorbed by $CO_2$ ice at the base of the slab. The bottom of the slab slowly warms up and $CO_2$ molecules begin to sublimate, leading to the accumulation of $CO_2$ gas underneath the slab. Eventually, the pressure exerted by the increasing volume of $CO_2$ gas exceeds the material strength of the overlying $CO_2$ ice, rupturing the slab. The trapped $CO_2$ gas escapes in high velocity jets, bringing along entrained regolith grains originating near the base of the slab. Some of this material falls back onto the ruptured slab, forming a thin layer of ejected regolith grains. The material falling back onto the $CO_2$ ice slab is mostly composed of tiny $H_2O$ ice grains, with similar spectral properties to regolith material elsewhere on these moons.

A possibly analogous solid-state greenhouse warming process occurs in “Cryptic” terrains near the south pole of Mars. Here, slabs of $CO_2$ ice, built up during the cold Martian winter, are ruptured by high velocity $CO_2$ jets during spring, which deposit entrained regolith material on top of the $CO_2$ slabs (*e.g.*, Kieffer et al., 2000, 2006, Chinnery et al., 2018). A similar solid-state, solar-driven process has been suggested to explain the plume activity and dark streaks observed on Triton during the Voyager 2 flyby of the Neptunian system (*e.g.*, Smith et al., 1989). On Triton, sunlight could pass through a translucent upper layer of nitrogen ice and interact with a dark absorbing layer beneath, which warms the base of the overlying nitrogen ice slab, driving the buildup and subsequent eruption of nitrogen-rich gas with entrained dark material (*e.g.*, Matson and Brown, 1989; Soderblom et al., 1990).

It is uncertain whether the base of $CO_2$ ice slabs on the Uranian moons could reach temperatures high enough to drive subsurface sublimation of $CO_2$, nor whether this process would occur fast enough to promote $CO_2$ gas accumulation, as opposed to $CO_2$ gas slowly escaping through fractures and other conduits to their surfaces. Voyager 2/ISS did not detect evidence for expansive slabs of $CO_2$ ice on Ariel, possibly because the spatial resolution of the ISS images is too low (~1 km/pixel; Smith et al., 1986) to detect these deposits, if they are present. Furthermore, Voyager 2/ISS did not detect plume activity or dark streaks on Ariel like those observed on Triton. Nevertheless, in theory a solid-state greenhouse process could help explain why the spectral signature of $CO_2$ ice appears to be obscured by tiny $H_2O$ ice grains on Ariel. Subsequent numerical modeling work of solid-state greenhouse warming on the Uranian moons, which is far beyond the scope of this paper, is needed to investigate this possible mechanism.

## 5. Summary

We analyzed new IRAC Ch.1 and Ch.2 geometric albedos to test the hypothesis that the surfaces of the classical Uranian satellites are mantled by tiny $H_2O$ ice grains. Radiative transfer modeling of these new data support the ubiquitous presence of tiny $H_2O$ ice grains (≤ 2 µm diameters), consistent with previous analyses of other long NIR datasets (~3 – 5 µm; Cartwright et al., 2015, 2018). Furthermore, our results indicate that these moons have higher IRAC albedos on their trailing hemispheres compared to their leading hemispheres (except for Miranda), suggesting larger abundances of tiny $H_2O$ ice grains and/or less $H_2O$ ice is exposed on their trailing sides.

We explored possible mechanisms to explain the apparent leading/trailing asymmetry in these moons' IRAC albedos and $H_2O$ ice band strengths, finding that the most likely driver is the presence of intimately mixed $CO_2$ ice on their trailing hemispheres. The absence of $CO_2$ ice on Miranda can also explain why this moon does not display hemispherical asymmetries in its $H_2O$ ice band strengths and IRAC albedos. Additionally, our results uncovered five regional-scale albedo zones on Ariel, which likely stem from the distribution of $CO_2$ ice on this moon. We explored the possible roles of micrometeorite impacts and volatile outgassing on the microstructure of these moons' regoliths, finding that both processes could contribute to developing the porous upper layers suggested by VIS polarimetric observations and our long NIR spectral models. We also discussed a possible mechanism for obscuring the spectral signature of areally mixed $CO_2$ ice via jetting of $CO_2$ gas and entrained regolith material.

This study builds upon previous efforts that investigated how the distribution of constituents on the classical Uranian satellites compares to icy bodies elsewhere. Icy satellites in the Jovian and Saturnian systems have relatively young and warm surfaces that are exposed to larger fluxes of charged particles compared to the Uranian moons. Consequently, the combination of geologic, thermal, and radiolytic processes operating on the surfaces of the Jovian and Saturnian moons is likely more efficient at removing tiny $H_2O$ grains compared to the relatively cold and quiescent environment of the Uranian system. Complementing our findings reported here, VIS polarimetry indicates that the Uranian moons have steep negative polarization branches (NPBs), consistent with porous upper layers dominated by tiny grains (Afanasiev et al., 2014). The Uranian moons' steep NPBs are comparable to the NPBs of similarly sized, $H_2O$-bearing TNOs, which could be mantled by porous upper layers developed by substantial volatile outgassing. Consequently, the surfaces of the Uranian satellites have different spectral properties than Jovian and Saturnian icy moons, and instead, perhaps the spectral signature of their surfaces are more similar to mid-sized, $H_2O$ ice-bearing TNOs like Charon, Orcus, Varuna, Ixion, Huya, and 1999 DE9.

## 6. Future Work

High signal-to-noise (S/N) spectral observations of the Uranian satellites at wavelengths > 2.5 µm are critical to investigate the prevalence of tiny $H_2O$ grains on their surfaces and the processes shaping the microstructure of their regoliths. Observations by the NIRSpec and MIRI spectrographs onboard the James Webb Space Telescope could dramatically expand our understanding of these moons across a wide swath of wavelengths (~0.6 – 29.5 µm). Observations made by next generation space telescopes, such as the proposed LUVOIR mission (*e.g.*, Bolcar et al., 2017), could collect spatially resolved and high S/N spectra of these moons, providing invaluable information about their spectral properties across the 1 to 5 µm range (Cartwright et al., 2019a). The Extremely Large Telescopes (ELTs), coming online over the next decade, could also provide spatially resolved, high-quality spectra of these moons over short NIR wavelengths (~1 – 2.5 µm) (Wong et al. 2019). Additionally, a spacecraft mission to the Uranian system would achieve far superior spatial resolution compared to any existing or proposed ground- or space-based facility (Cartwright et al., 2019b), and would therefore provide the best possible data for mapping the distribution of constituents and characterizing geologic landforms on these moons. Such a mission, returning *in situ* data, has the promise to revolutionize our understanding of ice giant planetary systems, which may represent a common type of exoplanetary system.

Theoretical work and physical experiments are also sorely needed to improve our understanding of the Uranian satellites' surface environments. Numerical models that investigate interactions between Uranus' magnetosphere and the classical moons are needed to determine the longitudes and latitudes of peak irradiation, and whether charged particle fluxes are sufficient to drive a radiolytic production cycle of $CO_2$ and perhaps other oxidized species like carbonic acid ($H_2CO_3$), carbon suboxide ($C_3O_2$), methanol ($CH_3OH$), and formaldehyde ($H_2CO$) (*e.g.*, Delitsky and Lane, 1997). Dynamical modeling of µ-ring particles on slowly expanding orbits that cross into the orbital zone of Miranda, and updated estimates of the rate of regolith overturn spurred by heliocentric micrometeorites on Miranda, are needed to investigate how these different sources of dust modify its surface composition. Thermodynamical modeling work is needed to explore whether thick slabs of $CO_2$ ice could undergo solid state greenhouse warming at cryogenic temperatures relevant to the Uranian system. Development and proliferation of new radiative transfer modeling codes that can account for distinct compositional layers would greatly improve our understanding of the regolith microstructure of the Uranian satellites and other icy objects. Similarly, cryogenic laboratory work that investigates the spectral and polarimetric properties of different particulate substrates, with stratified compositions and grain sizes, would provide new key knowledge of how layered media influence the spectral signature of icy object regoliths throughout the outer Solar System.

## Acknowledgements

This study was funded by a NASA Earth and Space Science Fellowship (grant number NNX14AP16H), as well as NASA Solar System Observing grant 16-SSO016_2-0070. We wish to recognize and acknowledge the significant cultural role and reverence of the summit of Mauna Kea within the indigenous Hawaiian community and to express our appreciation for the opportunity to observe from this special mountain. We thank the IRTF telescope operators and staff for providing observing support. Correspondence with Tom A. Nordheim and Michael M. Sori helped inform parts of the discussion section. We also thank Michael M. Sori and an anonymous reviewer for providing insightful feedback.

## 7. References

Afanasiev, V.L., Rosenbush, V.K. and Kiselev, N.N., 2014. Polarimetry of major Uranian moons at the 6-m telescope. *Astrophysical Bulletin 69* (2), p.211.

Bagnulo, S., Belskaya, I., Muinonen, K., Tozzi, G.P., Barucci, M.A., Kolokolova, L. and Fornasier, S., 2008. Discovery of two distinct polarimetric behaviours of trans-Neptunian objects. *Astronomy & Astrophysics 491* (2), p.L33.

Bagnulo, S., Belskaya, I., Boehnhardt, H., Kolokolova, L., Muinonen, K., Sterzik, M. and Tozzi, G.P., 2011. Polarimetry of small bodies of the solar system with large telescopes. *Journal of Quantitative Spectroscopy and Radiative Transfer 112* (13), p.2059.

Bauer, J.M., Roush, T.L., Geballe, T.R., Meech, K.J., Owen, T.C., Vacca, W.D., Rayner, J.T. and Jim, K.T., 2002. The near infrared spectrum of Miranda: Evidence of crystalline water ice. *Icarus*, *158* (1), p.178.

Beddingfield, C.B., Burr, D.M., Emery, J.P., 2015. Fault geometries on Uranus' satellite Miranda: Implications for internal structure and heat flow. *Icarus 247*, 35. http://dx.doi.org/10.1016/j.icarus.2014.09.048.

Beddingfield, C.B. and Cartwright, R.J. [*Under Review*]. Hidden tectonism on Miranda's Elsinore Corona revealed by polygonal impact craters.

Bell III, J.F. and McCord, T.B., 1991. A search for spectral units on the Uranian satellites using color ratio images. In *Lunar and Planetary Science Conference Proceedings*, Vol. 21, p. 473.

Belskaya, I.N., Bagnulo, S., Stinson, A., Tozzi, G.P., Muinonen, K., Shkuratov, Y.G., Barucci, M.A. and Fornasier, S., 2012. Polarimetry of trans-Neptunian objects (136472) Makemake and (90482) Orcus. *Astronomy & Astrophysics 547*, p.A101.

Bennett, C.J., Pirim, C. and Orlando, T.M., 2013. Space-weathering of solar system bodies: a laboratory perspective. *Chemical reviews*, *113* (12), p.9086.

Bevington, P.R. and Robinson, D.K., 1969. Testing the Fit. *Data Reduction and Error Analysis for the Physical Sciences*, p. 194.

Bohren, C.F. and Huffman, D.R., 1983. Absorption and scattering by a sphere. *Absorption and Scattering of Light by Small Particles*, p.82.

Bolcar, M.R., Aloezos, S., Bly, V.T., Collins, C., Crooke, J., Dressing, C.D., Fantano, L., Feinberg, L.D., France, K., Gochar, G. and Gong, Q., 2017, September. The large uv/optical/infrared surveyor (luvoir): Decadal mission concept design update. In *UV/Optical/IR Space Telescopes and Instruments: Innovative Technologies and Concepts VIII* (Vol. 10398, p. 1039809). International Society for Optics and Photonics.

Boxe, C.S., Bodsgard, B.R., Smythe, W. and Leu, M.T., 2007. Grain sizes, surface areas, and porosities of vapor-deposited H2O ices used to simulate planetary icy surfaces. *Journal of colloid and interface science 309* (2), p.412.

Brown, R.H. and Clark, R.N., 1984. Surface of Miranda: Identification of water ice. *Icarus 58* (2), p.288.

Brown, R.H. and Cruikshank, D.P., 1983. The Uranian satellites: Surface compositions and opposition brightness surges. *Icarus 55* (1), p.83.

Brown, M.E., 2012. The compositions of Kuiper belt objects. *Annual Review of Earth and Planetary Sciences*, *40*, pp.467-494.

Buratti, B.J. and Mosher, J.A., 1991. Comparative global albedo and color maps of the uranian satellites. *Icarus 90* (1), p.1.

Cartwright, R.J., Emery, J.P., Rivkin, A.S., Trilling, D.E. and Pinilla-Alonso, N., 2015. Distribution of $CO_2$ ice on the large moons of Uranus and evidence for compositional stratification of their near surfaces. *Icarus 257*, p. 428.

Cartwright, R.J., Emery, J.P., Pinilla-Alonso, N., Lucas, M.P., Rivkin, A.S., and Trilling, D.E., 2018. Red material on the large moons of Uranus: Dust from the irregular satellites? Icarus 314, 210-231.

Cartwright, R.J., Holler, B., Benecchi, S., Juanola-Parramon, R., Arney, G., Roberge, A. and Hammel, H., 2019a. Exploring the composition of icy bodies at the fringes of the Solar System with next generation space telescopes. *arXiv preprint arXiv:1903.07691*.

Cartwright, R.J., Emery, J.P., Pinilla-Alonso, N., Grundy, W.M., Cruikshank, D.P., 2019b. Probing the regoliths of the classical Uranian satellites using near-infrared telescope observations: Thick deposits of $CO_2$ ice mantled by a thin veneer of tiny $H_2O$ ice grains? In *AGU Fall Meeting Abstracts*.

Cassidy, T.A., Paranicas, C.P., Shirley, J.H., Dalton III, J.B., Teolis, B.D., Johnson, R.E., Kamp, L. and Hendrix, A.R., 2013. Magnetospheric ion sputtering and water ice grain size at Europa. *Planetary and Space Science 77*, p.64.

Chakarov, D.V., Gleeson, M.A. and Kasemo, B., 2001. Photoreactions of water and carbon at 90 K. *The Journal of Chemical Physics 115* (20), p.9477.

Chinnery, H.E., Hagermann, A., Kaufmann, E. and Lewis, S.R., 2018. The penetration of solar radiation into carbon dioxide ice. *Journal of Geophysical Research: Planet, 123* (4), p.864.

Clark, R.N., Carlson, R., Grundy, W. and Noll, K., 2013. Observed ices in the Solar System. In *The Science of Solar System Ices*, p.3. Springer New York.

Clark, R.N. and Lucey, P.G., 1984. Spectral properties of ice-particulate mixtures and implications for remote sensing: 1. Intimate mixtures. *Journal of Geophysical Research: Solid Earth*, *89* (B7), p.6341.

Clark, R.N. and Roush, T.L., 1984. Reflectance spectroscopy: Quantitative analysis techniques for remote sensing applications. *Journal of Geophysical Research: Solid Earth 89* (B7), p. 6329.

Croft, S.K. and Soderblom, L.A., 1991. Geology of the Uranian satellites. *Uranus*, p. 561.

Cruikshank, D.P., Morrison, D. and Pilcher, C.B., 1977. Identification of a new class of satellites in the outer solar system. *The Astrophysical Journal 217*, p.1006.

Cruikshank, D.P., 1980. Near-infrared studies of the satellites of Saturn and Uranus. *Icarus 41* (2), p. 246.

Cruikshank, D.P. and Brown, R.H., 1981. The uranian satellites: Water ice on Ariel and Umbriel. *Icarus 45* (3), p.607.

De Bergh, C., Schaller, E.L., Brown, M.E., Brunetto, R., Cruikshank, D.P. and Schmitt, B., 2013. The ices on transneptunian objects and Centaurs. In *The Science of Solar System Ices*. p. 107-146. Springer, New York, NY.

Delitsky, M.L., and Lane, A.L., 1997. Chemical schemes for surface modification of icy satellites: A road map. J. Geophys. Res. 102, 16385-16390.

De Pater, I., Hammel, H.B., Gibbard, S.G. and Showalter, M.R., 2006. New dust belts of Uranus: one ring, two ring, red ring, blue ring. *Science 312* (5770), p.92.

Emery, J.P., Cruikshank, D.P. and Van Cleve, J., 2006. Thermal emission spectroscopy (5.2–38 μm) of three Trojan asteroids with the Spitzer Space Telescope: Detection of fine-grained silicates. *Icarus 182* (2), p.496.

Fazio, G.G., Hora, J.L., Allen, L.E., Ashby, M.L.N., Barmby, P., Deutsch, L.K., Huang, J.S., Kleiner, S., Marengo, M., Megeath, S.T. and Melnick, G.J., 2004. The infrared array camera (IRAC) for the spitzer space telescope. *The Astrophysical Journal Supplement Series 154* (1), p.10.

Filacchione, G., Capaccioni, F., Ciarniello, M., Clark, R.N., Cuzzi, J.N., Nicholson, P.D., Cruikshank, D.P., Hedman, M.M., Buratti, B.J., Lunine, J.I. and Soderblom, L.A., 2012. Saturn's icy satellites and rings investigated by Cassini–VIMS: III–Radial compositional variability. *Icarus 220* (2), p.1064.

Gerakines, P.A. and Moore, M.H., 2001. Carbon suboxide in astrophysical ice analogs. *Icarus 154*, p.372.

Gerakines, P.A., Bray, J.J., Davis, A. and Richey, C.R., 2005. The strengths of near-infrared absorption features relevant to interstellar and planetary ices. *The Astrophysical Journal 620* (2), p.1140.

Gomis, O. and Strazzulla, G., 2005. CO2 production by ion irradiation of H2O ice on top of carbonaceous materials and its relevance to the Galilean satellites. *Icarus 177* (2), p.570-576.

Gourgeot, F., Dumas, C., Merlin, F., Vernazza, P. and Alvarez-Candal, A., 2014. Near-infrared spectroscopy of Miranda. *Astronomy & Astrophysics 562*, p. A46.

Grundy, W.M., Young, L.A. and Young, E.F., 2003. Discovery of $CO_2$ ice and leading–trailing spectral asymmetry on the uranian satellite ariel. *Icarus*, *162* (1), p.222-229.

Grundy, W.M., Young, L.A., Spencer, J.R., Johnson, R.E., Young, E.F. and Buie, M.W., 2006. Distributions of H2O and CO2 ices on Ariel, Umbriel, Titania, and Oberon from IRTF/SpeX observations. *Icarus 184* (2), p.543-555.

Grundy, W.M., Binzel, R.P., Buratti, B.J., Cook, J.C., Cruikshank, D.P., Dalle Ore, C.M., Earle, A.M., Ennico, K., Howett, C.J.A., Lunsford, A.W. and Olkin, C.B., 2016. Surface compositions across Pluto and Charon. *Science*, *351* (6279), p.9189.

Hanel, R., Conrath, B., Flasar, F.M., Kunde, V., Maguire, W., Pearl, J., Pirraglia, J., Samuelson, R., Cruikshank, D., Gautier, D. and Gierasch, P., 1986. Infrared observations of the Uranian system. *Science 233* (4759), p.70.

Hansen, G.B., 1997. Spectral absorption of solid CO2 from the ultraviolet to the far-infrared. *Advances in Space Research 20* (8), p.1613.

Hansen, G.B., 2005. Ultraviolet to near-infrared absorption spectrum of carbon dioxide ice from 0.174 to 1.8 μm. *Journal of Geophysical Research: Planets 110* (E11).

Hapke, 2012. Theory of Reflectance and Emittance Spectroscopy ($2^{nd}$ Ed.), p.404. Cambridge University Press.

Helfenstein, P., Thomas, P.C., Veverka, J., 1989. Evidence from Voyager II photometry for early resurfacing of Umbriel. *Nature 338*, p.324.

Helfenstein, P. et al., 1991. Oberon color photometry from Voyager and its geological implications. *Icarus 90*, p.14–29.

Hudson, R.L. and Moore, M.H., 2001. Radiation chemical alterations in solar system ices: an overview. *Journal of Geophysical Research: Planets 106* (E12), p.33275.

Jamieson, C.S., Mebel, A.M. and Kaiser, R.I., 2006. Understanding the kinetics and dynamics of radiation-induced reaction pathways in carbon monoxide ice at 10 K. *The Astrophysical Journal Supplement Series 163* (1), p.184.

Karkoschka, E., 2001. Comprehensive photometry of the rings and 16 satellites of Uranus with the Hubble Space Telescope. *Icarus 151*, p.51.

Kieffer, H.H., 2000, August. Annual punctuated CO2 slab-ice and jets on Mars. In *Second International Conference on Mars Polar Science and Exploration*, p.93.

Kieffer, H.H., Christensen, P.R. and Titus, T.N., 2006. CO2 jets formed by sublimation beneath translucent slab ice in Mars' seasonal south polar ice cap. *Nature 442* (7104), p.793.

Kim, Y.S. and Kaiser, R.I., 2012. Electron irradiation of Kuiper belt surface ices: Ternary N2-CH4-CO mixtures as a case study. *The Astrophysical Journal 758* (1), p.37.

Lewis, J.S., 1972. Low temperature condensation from the solar nebula. *Icarus 16* (2), p.241.

Lewis, J.S., 1973. Chemistry of the outer solar system. *Space Science Reviews 14* (3-4), p.401.

Loeffler, M.J., Baratta, G.A., Palumbo, M.E., Strazzulla, G. and Baragiola, R.A., 2005. CO synthesis in solid CO by Lyman-α photons and 200 keV protons. *Astronomy & Astrophysics 435* (2), p.587.

Mastrapa, R.M., Bernstein, M.P., Sandford, S.A., Roush, T.L., Cruikshank, D.P. and Dalle Ore, C.M., 2008. Optical constants of amorphous and crystalline $H_2O$ ice in the near infrared from 1.1 to 2.6 μm. *Icarus 197* (1), p.307.

Mastrapa, R.M., Sandford, S.A., Roush, T.L., Cruikshank, D.P. and Dalle Ore, C.M., 2009. Optical Constants of Amorphous and Crystalline $H_2O$ ice: 2.5-22 μm (4000-455 $cm^{-1}$) Optical Constants of $H_2O$ ice. *The Astrophysical Journal 701* (2), p.1347.

Matson, D.L. and Brown, R.H., 1989. Solid-state greenhouse and their implications for icy satellites. *Icarus 77* (1), p.67.

Mennella, V., Palumbo, M.E. and Baratta, G.A., 2004. Formation of CO and CO2 molecules by ion irradiation of water ice-covered hydrogenated carbon grains. *The Astrophysical Journal 615* (2), p.1073.

Mennella, V., Baratta, G.A., Palumbo, M.E. and Bergin, E.A., 2006. Synthesis of CO and CO2 molecules by UV irradiation of water ice-covered hydrogenated carbon grains. *The Astrophysical Journal 643* (2), p.923.

Moersch, J.E. and Christensen, P.R., 1995. Thermal emission from particulate surfaces: A comparison of scattering models with measured spectra. *Journal of Geophysical Research: Planets 100* (E4), p.7465.

Ness, N.F., et al., 1986. Magnetic fields at Uranus. *Science 233*, p.85-90.

Pappalardo, R.T., Reynolds, S.J., Greeley, R., 1997. Extensional tilt blocks on Miranda: Evidence for an upwelling origin of Arden Corona. *Journal Geophysical Research: Planets* 102 (E6), p.13369.

Poch, O., Cerubini, R., Pommerol, A., Jost, B. and Thomas, N., 2018. Polarimetry of water ice particles providing insights on grain size and degree of sintering on icy planetary surfaces. *Journal of Geophysical Research: Planets 123* (10), p.2564.

Prinn, R.G. and Fegley Jr, B., 1981. Kinetic inhibition of CO and N2 reduction in circumplanetary nebulae-Implications for satellite composition. *The Astrophysical Journal 249*, p.308-317.

Prinn, R.G. and Fegley Jr, B., 1989. Origin and Evolution of Planetary and Satellite Atmospheres. *SK Atreya, JB Pollack, MS Matthews, Eds*, p.78.

Raut, U., Fulvio, D., Loeffler, M.J. and Baragiola, R.A., 2012. Radiation synthesis of carbon dioxide in ice-coated carbon: implications for interstellar grains and icy moons. *The Astrophysical Journal 752* (2), p.159.

Rayner, J.T. et al., 1998. SpeX: A medium-resolution IR spectrograph for IRTF. *Proceedings of SPIE 3354*, p.468.

Rayner, J.T. et al., 2003. SpeX: A medium-resolution 0.8–5.5 micron spectrograph and imager for the NASA Infrared Telescope Facility. *Astronomical Society of the Pacific 115*, p.362.

Reach, W.T., Megeath, S.T., Cohen, M., Hora, J., Carey, S., Surace, J., Willner, S.P., Barmby, P., Wilson, G., Glaccum, W. and Lowrance, P., 2005. Absolute calibration of the infrared array camera on the spitzer space telescope. *Publications of the Astronomical Society of the Pacific 117* (835), p.978.

Schaller, E.L. and Brown, M.E., 2007. Volatile loss and retention on Kuiper belt objects. *The Astrophysical Journal Letters 659* (1), p.L61.

Sedlacko, T., Balog, R., Lafosse, A., Stano, M., Matejcik, S., Azria, R. and Illenberger, E., 2005. Reactions in condensed formic acid (HCOOH) induced by low energy (< 20 eV) electrons. *Physical Chemistry Chemical Physics 7* (6), p.1277.

Showalter, M.R. and Lissauer, J.J., 2006. The second ring-moon system of Uranus: discovery and dynamics. *Science 311* (5763), p.973.

Smith, B.A., Soderblom, L., Beebe, R., Boyce, J., Briggs, G., Bunker, A., Collins, S.A., Hansen, C.J., Johnson, T.V., Mitchell, J.L. and Terrile, R.J., 1981. Encounter with Saturn: Voyager 1 imaging science results. *Science 212* (4491), p.163.

Smith, B.A., Soderblom, L., Batson, R., Bridges, P., Inge, J.A.Y., Masursky, H., Shoemaker, E., Beebe, R., Boyce, J., Briggs, G. and Bunker, A., 1982. A new look at the Saturn system: The Voyager 2 images. *Science 215* (4532), p.504.

Smith, B.A. et al., 1986. Voyager 2 in the uranian system: Imaging science results. *Science 233*, p.43.

Smith, B.A., Soderblom, L.A., Banfield, D., Basilevsky, A.T., Beebe, R.F., Bollinger, K., Boyce, J.M., Brahic, A., Briggs, G.A., Brown, R.H. and Colvin, T., 1989. Voyager 2 at Neptune: Imaging science results. *Science*, *246* (4936), p.1422.

Soderblom, L.A., Kieffer, S.W., Becker, T.L., Brown, R.H., Cook, A.F., Hansen, C.J., Johnson, T.V., Kirk, R.L. and Shoemaker, E.M., 1990. Triton's geyser-like plumes: Discovery and basic characterization. *Science*, *250* (4979), pp.410.

Soifer, B.T., Neugebauer, G. and Matthews, K., 1981. Near-infrared spectrophotometry of the satellites and rings of Uranus. *Icarus 45* (3), p.612.

Sori, M.M., Bapst, J., Bramson, A.M., Byrne, S. and Landis, M.E., 2017. A Wunda-full world? Carbon dioxide ice deposits on Umbriel and other Uranian moons. *Icarus 290*, p.1.

Spiegel, M.R., 1992. Theory and Problems of Probability and Statistics. McGraw-Hill, New York, p.117.

Stone, E.C., Miller, E.D., 1986. The Voyager-2 encounter with the uranian system. *Science 233*, p.39.

Stone, E.C., Cooper, J.F., Cummings, A.C., McDonald, F.B., Trainor, J.H., Lal, N., McGuire, R. and Chenette, D.L., 1986. Energetic charged particles in the Uranian magnetosphere. *Science 233* (4759), p.93.

Strazzulla G, Palumbo M.E., 1998. Evolution of icy surfaces: an experimental approach. *Planetary Space Science 46*, p.1339.

Tamayo, D., Burns, J.A. and Hamilton, D.P., 2013. Chaotic dust dynamics and implications for the hemispherical color asymmetries of the Uranian satellites. *Icarus 226* (1), p. 655.

Taylor, J., 1997. Introduction to error analysis, the study of uncertainties in physical measurements (2nd Ed.). University Science Books, Sausalito, CA.

Werner, M.W., Roellig, T.L., Low, F.J., Rieke, G.H., Rieke, M., Hoffmann, W.F., Young, E., Houck, J.R., Brandl, B., Fazio, G.G. and Hora, J.L., 2004. The Spitzer space telescope mission. *The Astrophysical Journal Supplement Series 154* (1), p.1.

Wong, M.H., Cartwright, R., Chanover, N., Sayanagi, K., Greathouse, T., Tiscareno, M., Giles, R., Orton, G., Trilling, D., Sinclair, J. and Pinilla-Alonso, N., 2019. Solar system Deep Time-Surveys of atmospheres, surfaces, and rings. *arXiv preprint arXiv:1903.06321*.

Zahnle, K., Schenk, P., Levison, H. and Dones, L., 2003. Cratering rates in the outer Solar System. *Icarus 163* (2), p.263.

## Appendix A: Additional Methodology Details

### *A1. Uncertainty Estimation for Mean Albedos*

In this appendix, we describe our uncertainty estimation procedure for the mean IRAC albedos reported in section *3.2*. To do this, we summed the errors of each individual albedo measurement in quadrature [$\delta q$ = sqrt($\delta x^2 + \delta y^2 \ldots \delta i^2$)] and divided by the sample size ($n$) (*e.g.*, Taylor, 1997). Next, we calculated the standard deviation of the mean ($\sigma_{\bar{x}} = \sigma$/sqrt($n$)) to estimate point-to-point variation amongst the individual data points. We then added $\delta q/n$ and $\sigma_{\bar{x}}$ in quadrature, resulting in a final error for each mean albedo (reported in Table 4, Figures 2 and 3).

### *A2. Radiative Transfer Modeling*

Here, we provide additional description of the Hapke-Mie codes we utilized in this study. This hybrid spectral modeling approach calculates the single scattering albedo ($\varpi_0$) for each constituent using Mie theory (*e.g.*, Bohren and Huffman, 1983). These $\varpi_0$ values are then passed along to Hapke equations (*e.g.*, Hapke, 2012). Mie theory describes absorption and scattering by spherical particles of any grain size, which are isolated from each other by random distances. Although more computationally intensive, Mie codes provide a good approach for generating synthetic spectra that include grains similar in size and/or smaller than the wavelength of incident light, unlike "pure" Hapke models, which do not generate good results when considering tiny grains (*e.g.*, Moersch and Christensen, 1995; Emery et al., 2006). These Mie codes can introduce low amplitude resonance artifacts at different wavelengths, depending on the modeled grain size of each constituent. These artifacts are removed by using a narrow spread of diameters for each constituent (typically ~10% spread in grain size), which are averaged together into one grain size.

**Table 1:** The classical Uranian satellites

| Satellite | Orbital Radius (km) | Orbital Radius ($R_{Uranus}$) | Orbital Period (days) | Radius (km) | Mass (x $10^{20}$ kg) | Density (g $cm^{-3}$) | *Geo. Albedo ($A_0$) ($\lambda$ ~0.96 μm) |
|---|---|---|---|---|---|---|---|
| Miranda | 129,900 | 5.12 | 1.41 | 236 | 0.66 | 1.21 | 0.45 ± 0.02 |
| Ariel | 190,900 | 7.53 | 2.52 | 579 | 13.53 | 1.59 | 0.56 ± 0.02 |
| Umbriel | 266,000 | 10.5 | 4.14 | 585 | 11.72 | 1.46 | 0.26 ± 0.01 |
| Titania | 436,300 | 17.2 | 8.71 | 789 | 35.27 | 1.66 | 0.39 ± 0.02 |
| Oberon | 583,500 | 23.0 | 13.46 | 762 | 30.14 | 1.56 | 0.33 ± 0.01 |

**Geometric albedos from Fig. 7 in Karkoschka (2001).*

**Table 2:** Spitzer/IRAC observations (Program 71 and 11112)

| Target | Subsolar Long. (°) | Subsolar Lat. (°) | Prog. ID | Observation | UT Date | UT Time (mid-expos) | Number Exposures | $t_{int}$ per Channel (s) | Phase Angle | Heliocentric Distance (AU) | Observer Distance (AU) |
|---|---|---|---|---|---|---|---|---|---|---|---|
| Miranda | 78.35 | 32.19 | 11112 | Ariel L2 | 10/17/2015 | 18:02 | 10 | 268 | 2.926 | 19.982 | 19.863 |
| | 81.18 | 32.19 | 11112 | Titania T2 | 10/17/2015 | 18:18 | 10 | 268 | 2.926 | 19.982 | 19.863 |
| | 84.01 | 32.19 | 11112 | Oberon T2 | 10/17/2015 | 18:34 | 10 | 268 | 2.926 | 19.982 | 19.862 |
| | 253.51 | 32.19 | 11112 | Ariel L1 | 10/17/2015 | 0:37 | 10 | 268 | 2.928 | 19.982 | 19.875 |
| | 256.34 | 32.19 | 11112 | Titania T1 | 10/17/2015 | 0:53 | 10 | 268 | 2.928 | 19.982 | 19.874 |
| | 259.17 | 32.19 | 11112 | Oberon T1 | 10/17/2015 | 1:09 | 10 | 268 | 2.928 | 19.982 | 19.874 |
| Ariel | 3.2 | 33.8 | 11112 | Ariel T2 | 10/21/2015 | 3:23 | 10 | 268 | 2.907 | 19.983 | 19.807 |
| | 4.8 | 33.8 | 11112 | Umbriel T2 | 10/21/2015 | 3:39 | 10 | 268 | 2.907 | 19.983 | 19.807 |
| | 6.4 | 33.8 | 11112 | Titania L2 | 10/21/2015 | 3:55 | 10 | 268 | 2.907 | 19.983 | 19.807 |
| | 8.0 | 33.8 | 11112 | Oberon L1 | 10/21/2015 | 4:11 | 10 | 268 | 2.907 | 19.983 | 19.806 |
| | 54.0 | -18.4 | *71 | Titania_T | 12/3/2003 | 11:08 | 3 | 80 | 2.836 | 20.037 | 20.174 |
| | 82.0 | 33.9 | 11112 | Umbriel L1 | 10/26/2015 | 17:36 | 10 | 268 | 2.857 | 19.981 | 19.714 |
| | 87.8 | -6.7 | *71 | Ariel L | 6/10/2005 | 4:04 | 3 | 80 | 2.90 | 20.0660 | 20.0770 |
| | 92.8 | -10.7 | *71 | Umbriel L | 6/29/2004 | 21:55 | 3 | 80 | 2.57 | 20.0485 | 19.5529 |
| | 93.9 | -6.65 | *71 | Umbriel T | 6/15/2005 | 6:03 | 3 | 80 | 2.90 | 20.0661 | 19.9914 |
| | 106.5 | 33.8 | 11112 | Oberon L2 | 10/21/2015 | 20:44 | 10 | 268 | 2.903 | 19.981 | 19.794 |
| | 127.4 | -14.6 | *71 | Oberon L | 11/23/2004 | 8:16 | 3 | 80 | 2.86 | 20.0556 | 19.8728 |
| | 135.3 | 33.8 | 11112 | Ariel L1 | 10/17/2015 | 0:37 | 10 | 268 | 2.929 | 19.981 | 19.874 |
| | 136.9 | 33.8 | 11112 | Titania T1 | 10/17/2015 | 0:53 | 10 | 268 | 2.929 | 19.981 | 19.874 |
| | 138.5 | 33.8 | 11112 | Oberon T1 | 10/17/2015 | 1:09 | 10 | 268 | 2.929 | 19.981 | 19.874 |
| | 188.3 | 33.9 | 11112 | Umbriel L2 | 10/27/2015 | 11:27 | 10 | 268 | 2.849 | 19.980 | 19.700 |
| | 239.0 | 33.8 | 11112 | Ariel L2 | 10/17/2015 | 18:02 | 10 | 268 | 2.925 | 19.981 | 19.862 |
| | 240.6 | 33.8 | 11112 | Titania T2 | 10/17/2015 | 18:18 | 10 | 268 | 2.925 | 19.981 | 19.862 |
| | 242.2 | 33.8 | 11112 | Oberon T2 | 10/17/2015 | 18:34 | 10 | 268 | 2.925 | 19.981 | 19.862 |
| | 253.8 | 33.8 | 11112 | Ariel T1 | 10/20/2015 | 9:00 | 10 | 268 | 2.912 | 19.981 | 19.818 |
| | 255.4 | 33.8 | 11112 | Umbriel T1 | 10/20/2015 | 9:16 | 10 | 268 | 2.912 | 19.981 | 19.818 |
| | 255.5 | -18.4 | *71 | Oberon T | 12/2/2003 | 8:28 | 3 | 80 | 2.84 | 20.0361 | 20.1543 |
| | 257.0 | 33.8 | 11112 | Titania L1 | 10/20/2015 | 9:32 | 10 | 268 | 2.912 | 19.981 | 19.818 |
| | 276.5 | -14.6 | *71 | Ariel T | 11/26/2004 | 21:48 | 3 | 80 | 2.88 | 20.0566 | 19.9335 |
| Umbriel | 23.18 | 33.44 | 11112 | Ariel L2 | 10/17/2015 | 18:02 | 10 | 80 | 2.925 | 19.983 | 19.864 |
| | 24.15 | 33.44 | 11112 | Titania T2 | 10/17/2015 | 18:18 | 10 | 268 | 2.925 | 19.983 | 19.864 |
| | 25.11 | 33.44 | 11112 | Oberon T2 | 10/17/2015 | 18:34 | 10 | 268 | 2.925 | 19.983 | 19.864 |
| | 27.7 | 33.46 | 11112 | Oberon L2 | 10/21/2015 | 20:44 | 10 | 268 | 2.901 | 19.983 | 19.794 |
| | 34.7 | -10.5 | *71 | Titania L | 11/29/2005 | 9:13 | 3 | 80 | 2.81 | 20.0748 | 19.7977 |
| | 82.8 | -10.7 | *71 | Umbriel L | 6/29/2004 | 21:55 | 3 | 80 | 2.57 | 20.0488 | 19.5532 |
| | 83.5 | 33.47 | 11112 | Umbriel L1 | 10/26/2015 | 17:36 | 10 | 268 | 2.857 | 19.981 | 19.714 |
| | 148.11 | 33.47 | 11112 | Umbriel L2 | 10/27/2015 | 11:27 | 10 | 268 | 2.849 | 19.980 | 19.700 |
| | 149.9 | -18.3 | *71 | Oberon T | 12/2/2003 | 8:28 | 3 | 80 | 2.84 | 20.0349 | 20.1532 |
| | 151.8 | -14.5 | *71 | Ariel T | 11/26/2004 | 21:48 | 3 | 80 | 2.88 | 20.0550 | 19.9319 |
| | 202.3 | -14.5 | 71 | Oberon_L | 11/23/2004 | 8:16 | 3 | 80 | 2.862 | 20.056 | 19.873 |
| | 246.5 | -18.2 | *71 | Titania T | 12/3/2003 | 11:08 | 3 | 80 | 2.84 | 20.0358 | 20.1727 |
| | 251.12 | 33.46 | 11112 | Ariel T1 | 10/20/2015 | 9:00 | 10 | 268 | 2.912 | 19.981 | 19.818 |
| | 252.09 | 33.46 | 11112 | Umbriel T1 | 10/20/2015 | 9:16 | 10 | 268 | 2.912 | 19.981 | 19.818 |
| | 253.05 | 33.46 | 11112 | Titania L1 | 10/20/2015 | 9:32 | 10 | 268 | 2.912 | 19.981 | 19.818 |
| | 276.7 | -6.7 | *71 | Umbriel T | 6/15/2005 | 6:03 | 3 | 80 | 2.90 | 20.0663 | 19.9916 |
| | 317.66 | 33.46 | 11112 | Ariel T2 | 10/21/2015 | 3:23 | 10 | 268 | 2.907 | 19.983 | 19.807 |
| | 318.63 | 33.46 | 11112 | Umbriel T2 | 10/21/2015 | 3:39 | 10 | 268 | 2.907 | 19.983 | 19.807 |
| | 319.59 | 33.46 | 11112 | Titania L2 | 10/21/2015 | 3:55 | 10 | 268 | 2.906 | 19.983 | 19.807 |
| | 320.14 | 33.44 | 11112 | Ariel L1 | 10/17/2015 | 0:37 | 10 | 268 | 2.928 | 19.983 | 19.876 |
| | 320.56 | 33.46 | 11112 | Oberon L1 | 10/21/2015 | 4:11 | 10 | 268 | 2.906 | 19.983 | 19.807 |
| | 321.11 | 33.44 | 11112 | Titania T1 | 10/17/2015 | 0:53 | 10 | 268 | 2.928 | 19.983 | 19.876 |
| | 322.07 | 33.44 | 11112 | Oberon T1 | 10/17/2015 | 1:09 | 10 | 268 | 2.928 | 19.983 | 19.876 |
| Titania | 6.84 | -14.51 | 71 | Ariel T | 11/26/2004 | 21:48 | 3 | 80 | 2.88 | 20.0566 | 19.9335 |
| | 12.63 | -6.69 | 71 | Umbriel T | 6/15/2005 | 6:03 | 3 | 80 | 2.90 | 20.0663 | 19.9916 |
| | 51.05 | 33.42 | 11112 | Ariel T1 | 10/20/2015 | 9:00 | 10 | 268 | 2.912 | 19.983 | 19.820 |
| | 51.51 | 33.42 | 11112 | Umbriel T1 | 10/20/2015 | 9:16 | 10 | 268 | 2.912 | 19.983 | 19.820 |
| | 51.97 | 33.42 | 11112 | Titania L1 | 10/20/2015 | 9:32 | 10 | 268 | 2.912 | 19.983 | 19.820 |
| | 82.72 | 33.42 | 11112 | Ariel T2 | 10/21/2015 | 3:23 | 10 | 268 | 2.907 | 19.982 | 19.806 |
| | 83.1 | -10.5 | *71 | Titania L | 11/29/2005 | 9:13 | 3 | 80 | 2.81 | 20.0737 | 19.7966 |
| | 83.18 | 33.42 | 11112 | Umbriel T2 | 10/21/2015 | 3:39 | 10 | 268 | 2.907 | 19.982 | 19.806 |
| | 83.64 | 33.42 | 11112 | Titania L2 | 10/21/2015 | 3:55 | 10 | 268 | 2.907 | 19.982 | 19.806 |
| | 84.1 | 33.42 | 11112 | Oberon L1 | 10/21/2015 | 4:11 | 10 | 268 | 2.907 | 19.982 | 19.806 |
| | 112.59 | 33.43 | 11112 | Oberon L2 | 10/21/2015 | 20:44 | 10 | 268 | 2.903 | 19.981 | 19.793 |
| | 162.41 | -6.74 | 71 | Ariel_L | 6/10/2005 | 4:04 | 3 | 80 | 2.897 | 20.0632 | 20.0741 |
| | 216.7 | -18.3 | *71 | Oberon T | 12/2/2003 | 8:28 | 3 | 80 | 2.84 | 20.0341 | 20.1524 |
| | 219.5 | -14.5 | *71 | Oberon L | 11/23/2004 | 8:16 | 3 | 80 | 2.86 | 20.0541 | 19.8713 |
| | 262.6 | -18.3 | *71 | Titania T | 12/3/2003 | 11:08 | 3 | 80 | 2.84 | 20.0361 | 20.1729 |
| | 272.55 | 33.4 | 11112 | Ariel L1 | 10/17/2015 | 0:37 | 10 | 268 | 2.928 | 19.982 | 19.875 |
| | 273.01 | 33.4 | 11112 | Titania T1 | 10/17/2015 | 0:53 | 10 | 268 | 2.928 | 19.982 | 19.875 |
| | 273.47 | 33.4 | 11112 | Oberon T1 | 10/17/2015 | 1:09 | 10 | 268 | 2.928 | 19.982 | 19.874 |
| | 285.1 | -10.7 | *71 | Umbriel L | 6/29/2004 | 21:55 | 3 | 80 | 2.57 | 20.0492 | 19.5537 |
| | 302.56 | 33.4 | 11112 | Ariel L2 | 10/17/2015 | 18:02 | 10 | 268 | 2.925 | 19.983 | 19.864 |
| | 303.02 | 33.4 | 11112 | Titania T2 | 10/17/2015 | 18:18 | 10 | 268 | 2.925 | 19.983 | 19.864 |
| | 303.48 | 33.4 | 11112 | Oberon T2 | 10/17/2015 | 18:34 | 10 | 268 | 2.925 | 19.983 | 19.864 |
| | 314.01 | 33.43 | 11112 | Umbriel L1 | 10/26/2015 | 17:36 | 10 | 268 | 2.856 | 19.983 | 19.715 |
| | 344.76 | 33.43 | 11112 | Umbriel L2 | 10/27/2015 | 11:27 | 10 | 268 | 2.848 | 19.984 | 19.704 |
| Oberon | 39.74 | 33.26 | 11112 | Ariel T1 | 10/20/2015 | 9:00 | 10 | 268 | 2.912 | 19.984 | 19.821 |
| | 40.04 | 33.26 | 11112 | Umbriel T1 | 10/20/2015 | 9:16 | 10 | 268 | 2.912 | 19.984 | 19.821 |
| | 40.34 | 33.26 | 11112 | Titania L1 | 10/20/2015 | 9:32 | 10 | 268 | 2.912 | 19.984 | 19.821 |
| | 60.22 | 33.27 | 11112 | Ariel T2 | 10/21/2015 | 3:23 | 10 | 268 | 2.907 | 19.983 | 19.808 |
| | 60.52 | 33.27 | 11112 | Umbriel T2 | 10/21/2015 | 3:39 | 10 | 268 | 2.907 | 19.983 | 19.807 |
| | 60.82 | 33.27 | 11112 | Titania L2 | 10/21/2015 | 3:55 | 10 | 268 | 2.907 | 19.983 | 19.807 |
| | 61.11 | 33.27 | 11112 | Oberon L1 | 10/21/2015 | 4:11 | 10 | 268 | 2.907 | 19.983 | 19.807 |
| | 79.55 | 33.27 | 11112 | Oberon L2 | 10/21/2015 | 20:44 | 10 | 268 | 2.902 | 19.982 | 19.795 |
| | 88.3 | -14.6 | *71 | Oberon L | 11/23/2004 | 8:16 | 3 | 80 | 2.86 | 20.0564 | 19.8736 |
| | 133.5 | -10.7 | *71 | Umbriel L | 6/29/2004 | 21:55 | 3 | 80 | 2.57 | 20.0460 | 19.5504 |
| | 141.9 | -6.7 | *71 | Umbriel T | 6/15/2005 | 6:03 | 3 | 80 | 2.90 | 20.0632 | 19.9884 |
| | 183.6 | -14.6 | 71 | Ariel T | 11/26/2004 | 21:48 | 3 | 80 | 2.88 | 20.0566 | 19.9335 |
| | 209.79 | 33.29 | 11112 | Umbriel L1 | 10/26/2015 | 17:36 | 10 | 268 | 2.857 | 19.978 | 19.711 |
| | 229.68 | 33.29 | 11112 | Umbriel L2 | 10/27/2015 | 11:27 | 10 | 268 | 2.848 | 19.979 | 19.699 |

| | | | | | | | | | | |
|---|---|---|---|---|---|---|---|---|---|---|
| 261.9 | -18.3 | *71 | Oberon T | 12/2/2003 | 8:28 | 3 | 80 | 2.84 | 20.0358 | 20.1541 |
| 290.3 | -10.6 | *71 | Titania L | 11/29/2005 | 9:13 | 3 | 80 | 2.81 | 20.0747 | 19.7975 |
| 291.6 | -18.3 | *71 | Titania T | 12/3/2003 | 11:08 | 3 | 80 | 2.84 | 20.0378 | 20.1747 |
| 310.2 | 33.24 | 11112 | Ariel L1 | 10/17/2015 | 0:37 | 10 | 268 | 2.928 | 19.984 | 19.877 |
| 310.49 | 33.24 | 11112 | Titania T1 | 10/17/2015 | 0:53 | 10 | 268 | 2.928 | 19.984 | 19.877 |
| 310.79 | 33.24 | 11112 | Oberon T1 | 10/17/2015 | 1:09 | 10 | 268 | 2.928 | 19.984 | 19.876 |
| 329.6 | 33.25 | 11112 | Ariel L2 | 10/17/2015 | 18:02 | 10 | 268 | 2.925 | 19.985 | 19.865 |
| 329.9 | 33.25 | 11112 | Titania T2 | 10/17/2015 | 18:18 | 10 | 268 | 2.925 | 19.985 | 19.865 |
| 330.19 | 33.25 | 11112 | Oberon T2 | 10/17/2015 | 18:34 | 10 | 268 | 2.925 | 19.985 | 19.865 |

**Reported previously in Cartwright et al. (2015).*

**Table 3:** IRAC fluxes and albedos

| Target | Subsolar Long. (°) | Subsolar Lat. (°) | Prog. ID | Ch.1 Flux (μJy) | Ch.1 ΔFlux (μJy) | Ch.2 Flux (μJy) | Ch.2 ΔFlux (μJy) | Ch.1 Geo. Alb. | Ch.1 Geo. ΔAlb. | Ch.2 Geo. Alb. | Ch.2 Geo. ΔAlb. |
|---|---|---|---|---|---|---|---|---|---|---|---|
| Miranda | 78.35 | 32.19 | 11112 | 104.813 | 9.337 | - | - | 0.193 | 0.016 | - | - |
| | 81.18 | 32.19 | 11112 | 108.835 | 11.503 | - | - | 0.2 | 0.022 | - | - |
| | 84.01 | 32.19 | 11112 | 102.101 | 13.110 | - | - | 0.188 | 0.024 | - | - |
| | 253.51 | 32.19 | 11112 | 109.869 | 10.141 | - | - | 0.202 | 0.019 | - | - |
| | 256.34 | 32.19 | 11112 | 102.857 | 14.472 | - | - | 0.189 | 0.027 | - | - |
| | 259.17 | 32.19 | 11112 | 103.399 | 13.059 | - | - | 0.19 | 0.024 | - | - |
| Ariel | 3.2 | 33.8 | 11112 | 659.530 | 20.967 | - | - | 0.203 | 0.006 | - | - |
| | 4.8 | 33.8 | 11112 | 643.812 | 12.070 | - | - | 0.198 | 0.004 | - | - |
| | 6.4 | 33.8 | 11112 | 645.758 | 14.975 | - | - | 0.199 | 0.005 | - | - |
| | 8.0 | 33.8 | 11112 | 660.385 | 14.257 | - | - | 0.203 | 0.004 | - | - |
| | 54.0 | -18.4 | *71 | 594.817 | 15.045 | 212.256 | 21.882 | 0.191 | 0.005 | 0.104 | 0.011 |
| | 82.0 | 33.9 | 11112 | 588.626 | 15.610 | 183.182 | 21.663 | 0.179 | 0.005 | 0.085 | 0.01 |
| | 87.8 | -6.7 | *71 | 594.398 | 32.904 | 194.828 | 25.527 | 0.189 | 0.01 | 0.095 | 0.012 |
| | 92.8 | -10.7 | *71 | 620.646 | 18.584 | 187.376 | 16.891 | 0.185 | 0.006 | 0.085 | 0.008 |
| | 93.9 | -6.65 | *71 | 568.083 | 7.894 | 184.778 | 17.541 | 0.188 | 0.006 | 0.089 | 0.008 |
| | 106.5 | 33.8 | 11112 | 575.153 | 5.275 | 184.812 | 16.093 | 0.177 | 0.002 | 0.086 | 0.008 |
| | 127.4 | -14.6 | *71 | 596.511 | 283.830 | 186.831 | 0.025 | 0.186 | 0.009 | 0.089 | 0.012 |
| | 135.3 | 33.8 | 11112 | 572.080 | 8.669 | 172.272 | 17.264 | 0.177 | 0.003 | 0.081 | 0.008 |
| | 136.9 | 33.8 | 11112 | 577.803 | 8.546 | 171.934 | 19.100 | 0.179 | 0.006 | 0.082 | 0.01 |
| | 138.5 | 33.8 | 11112 | 561.731 | 10.040 | 173.325 | 20.385 | 0.174 | 0.003 | 0.082 | 0.01 |
| | 188.3 | 33.9 | 11112 | 676.394 | 15.659 | 193.205 | 20.102 | 0.206 | 0.005 | 0.089 | 0.009 |
| | 239.0 | 33.8 | 11112 | 700.484 | 6.856 | 227.880 | 12.688 | 0.217 | 0.002 | 0.107 | 0.006 |
| | 240.6 | 33.8 | 11112 | 696.190 | 6.407 | 226.929 | 13.424 | 0.216 | 0.002 | 0.107 | 0.006 |
| | 242.2 | 33.8 | 11112 | 684.961 | 8.316 | 231.291 | 12.768 | 0.212 | 0.003 | 0.109 | 0.006 |
| | 253.8 | 33.8 | 11112 | 736.567 | 22.843 | 273.547 | 7.795 | 0.227 | 0.007 | 0.128 | 0.004 |
| | 255.4 | 33.8 | 11112 | 720.757 | 13.907 | 260.801 | 13.981 | 0.230 | 0.004 | 0.127 | 0.007 |
| | 255.5 | -18.4 | *71 | 727.494 | 18.160 | 257.761 | 6.767 | 0.224 | 0.006 | 0.121 | 0.003 |
| | 257.0 | 33.8 | 11112 | 765.878 | 12.012 | 262.182 | 8.575 | 0.236 | 0.004 | 0.123 | 0.004 |
| | 276.5 | -14.6 | *71 | 743.609 | 15.803 | 253.504 | 17.660 | 0.233 | 0.005 | 0.121 | 0.008 |
| Umbriel | 23.18 | 33.44 | 11112 | 525.807 | 10.028 | 174.352 | 10.906 | 0.151 | 0.003 | 0.076 | 0.005 |
| | 24.15 | 33.44 | 11112 | 527.332 | 7.836 | 162.226 | 9.246 | 0.152 | 0.003 | 0.071 | 0.004 |
| | 25.11 | 33.44 | 11112 | 528.400 | 7.088 | 162.119 | 9.803 | 0.152 | 0.003 | 0.071 | 0.004 |
| | 27.7 | 33.46 | 11112 | 526.657 | 5.113 | 175.185 | 10.766 | 0.152 | 0.002 | 0.077 | 0.005 |
| | 34.7 | -10.5 | *71 | 542.207 | 10.337 | 198.499 | 20.268 | 0.155 | 0.003 | 0.087 | 0.009 |
| | 82.8 | -10.7 | *71 | 589.852 | 27.126 | 173.956 | 10.618 | 0.163 | 0.008 | 0.073 | 0.005 |
| | 83.5 | 33.47 | 11112 | 530.821 | 21.345 | 156.081 | 8.416 | 0.15 | 0.006 | 0.067 | 0.004 |
| | 148.11 | 33.47 | 11112 | 535.740 | 16.728 | 152.986 | 9.147 | 0.151 | 0.005 | 0.069 | 0.002 |
| | 149.9 | -18.3 | *71 | 501.836 | 10.500 | 171.094 | 13.820 | 0.149 | 0.003 | 0.077 | 0.006 |
| | 151.8 | -14.5 | *71 | 501.317 | 14.183 | 166.794 | 22.535 | 0.146 | 0.004 | 0.074 | 0.010 |
| | 202.3 | -14.5 | 71 | 485.058 | 8.162 | 159.173 | 4.837 | 0.140 | 0.003 | 0.070 | 0.002 |
| | 246.5 | -18.2 | *71 | 536.311 | 4.720 | 176.927 | 9.175 | 0.159 | 0.002 | 0.080 | 0.004 |
| | 251.12 | 33.46 | 11112 | 538.602 | 26.039 | 181.815 | 4.464 | 0.154 | 0.008 | 0.079 | 0.002 |
| | 252.09 | 33.46 | 11112 | 526.272 | 15.950 | 166.972 | 8.663 | 0.151 | 0.005 | 0.073 | 0.004 |
| | 253.05 | 33.46 | 11112 | 545.633 | 20.232 | 175.524 | 10.885 | 0.156 | 0.006 | 0.076 | 0.005 |
| | 276.7 | -6.7 | *71 | 535.371 | 6.137 | 166.121 | 7.682 | 0.157 | 0.002 | 0.074 | 0.004 |
| | 317.66 | 33.46 | 11112 | 535.962 | 9.524 | 171.661 | 9.967 | 0.153 | 0.003 | 0.075 | 0.004 |
| | 318.63 | 33.46 | 11112 | 532.655 | 9.305 | 172.045 | 11.052 | 0.152 | 0.003 | 0.075 | 0.005 |
| | 319.59 | 33.46 | 11112 | 545.655 | 9.583 | 166.764 | 10.197 | 0.156 | 0.003 | 0.073 | 0.004 |
| | 320.14 | 33.44 | 11112 | 533.055 | 7.628 | 162.144 | 11.205 | 0.154 | 0.003 | 0.071 | 0.005 |
| | 320.56 | 33.46 | 11112 | 539.198 | 10.413 | 169.466 | 12.361 | 0.155 | 0.003 | 0.074 | 0.005 |
| | 321.11 | 33.44 | 11112 | 537.943 | 8.604 | 166.479 | 10.706 | 0.155 | 0.003 | 0.073 | 0.005 |
| | 322.07 | 33.44 | 11112 | 534.798 | 7.971 | 161.950 | 9.724 | 0.154 | 0.003 | 0.071 | 0.004 |
| Titania | 6.84 | -14.51 | 71 | 995.886 | 8.666 | 286.429 | 11.853 | 0.161 | 0.002 | 0.071 | 0.003 |
| | 12.63 | -6.69 | 71 | 986.223 | 8.084 | 250.411 | 11.364 | 0.161 | 0.002 | 0.062 | 0.003 |
| | 51.05 | 33.42 | 11112 | 999.822 | 6.131 | 253.158 | 2.487 | 0.159 | 0.001 | 0.061 | 0.001 |
| | 51.51 | 33.42 | 11112 | 1000.129 | 8.841 | 255.826 | 2.425 | 0.159 | 0.002 | 0.062 | 0.001 |
| | 51.97 | 33.42 | 11112 | 997.358 | 7.949 | 255.885 | 2.369 | 0.159 | 0.001 | 0.062 | 0.001 |
| | 82.72 | 33.42 | 11112 | 981.652 | 9.228 | 246.710 | 3.690 | 0.156 | 0.002 | 0.06 | 0.001 |
| | 83.1 | -10.5 | *71 | 1001.106 | 7.886 | 261.483 | 9.439 | 0.160 | 0.001 | 0.063 | 0.002 |
| | 83.18 | 33.42 | 11112 | 995.596 | 10.004 | 252.702 | 2.524 | 0.158 | 0.002 | 0.061 | 0.001 |
| | 83.64 | 33.42 | 11112 | 995.593 | 9.885 | 255.187 | 1.987 | 0.158 | 0.002 | 0.061 | 0.001 |
| | 84.1 | 33.42 | 11112 | 986.474 | 11.418 | 250.819 | 3.172 | 0.157 | 0.002 | 0.061 | 0.001 |
| | 112.59 | 33.43 | 11112 | 997.846 | 5.781 | 248.301 | 2.154 | 0.158 | 0.001 | 0.06 | 0.001 |
| | 162.41 | -6.74 | 71 | 988.799 | 6.961 | 251.218 | 47.894 | 0.162 | 0.001 | 0.063 | 0.012 |
| | 216.7 | -18.3 | *71 | 955.511 | 9.567 | 263.413 | 7.501 | 0.157 | 0.002 | 0.066 | 0.002 |
| | 219.5 | -14.5 | *71 | 981.870 | 4.453 | 263.375 | 4.167 | 0.158 | 0.001 | 0.064 | 0.001 |
| | 262.6 | -18.3 | *71 | 982.228 | 11.316 | 264.992 | 3.861 | 0.162 | 0.002 | 0.067 | 0.001 |
| | 272.55 | 33.4 | 11112 | 1018.927 | 7.179 | 269.660 | 3.764 | 0.163 | 0.001 | 0.066 | 0.001 |
| | 273.01 | 33.4 | 11112 | 1012.587 | 5.315 | 265.873 | 2.507 | 0.162 | 0.001 | 0.065 | 0.001 |
| | 273.47 | 33.4 | 11112 | 999.674 | 6.991 | 272.157 | 5.215 | 0.160 | 0.001 | 0.066 | 0.001 |
| | 285.1 | -10.7 | *71 | 1062.645 | 5.235 | 279.446 | 6.087 | 0.163 | 0.001 | 0.065 | 0.001 |
| | 302.56 | 33.4 | 11112 | 1021.862 | 14.417 | 275.479 | 4.880 | 0.163 | 0.002 | 0.067 | 0.001 |
| | 303.02 | 33.4 | 11112 | 994.541 | 6.268 | 269.978 | 3.545 | 0.159 | 0.001 | 0.066 | 0.001 |
| | 303.48 | 33.4 | 11112 | 1003.458 | 8.370 | 271.028 | 4.947 | 0.16 | 0.002 | 0.066 | 0.001 |
| | 314.01 | 33.43 | 11112 | 1021.441 | 5.842 | 273.213 | 2.671 | 0.16 | 0.001 | 0.066 | 0.001 |
| | 344.76 | 33.43 | 11112 | 1041.993 | 9.302 | 278.576 | 6.945 | 0.163 | 0.002 | 0.066 | 0.001 |
| Oberon | 39.74 | 33.26 | 11112 | 971.386 | 7.088 | 269.102 | 2.488 | 0.166 | 0.002 | 0.07 | 0.001 |
| | 40.04 | 33.26 | 11112 | 961.566 | 7.140 | 262.602 | 2.564 | 0.164 | 0.002 | 0.068 | 0.001 |
| | 40.34 | 33.26 | 11112 | 970.823 | 7.312 | 264.107 | 3.681 | 0.166 | 0.002 | 0.069 | 0.001 |
| | 60.22 | 33.27 | 11112 | 955.480 | 5.914 | 256.168 | 4.889 | 0.163 | 0.002 | 0.066 | 0.001 |
| | 60.52 | 33.27 | 11112 | 961.027 | 8.846 | 256.649 | 2.940 | 0.164 | 0.002 | 0.067 | 0.001 |
| | 60.82 | 33.27 | 11112 | 962.733 | 4.424 | 253.901 | 4.799 | 0.164 | 0.001 | 0.066 | 0.001 |
| | 61.11 | 33.27 | 11112 | 969.298 | 8.106 | 255.176 | 2.931 | 0.165 | 0.002 | 0.066 | 0.001 |
| | 79.55 | 33.27 | 11112 | 962.301 | 4.702 | 257.660 | 2.189 | 0.164 | 0.001 | 0.067 | 0.001 |
| | 88.3 | -14.6 | *71 | 967.238 | 5.539 | 254.584 | 6.268 | 0.167 | 0.001 | 0.067 | 0.002 |
| | 133.5 | -10.7 | *71 | 1023.989 | 5.203 | 268.775 | 4.326 | 0.169 | 0.001 | 0.067 | 0.001 |
| | 141.9 | -6.7 | *71 | 939.477 | 5.427 | 254.564 | 2.914 | 0.164 | 0.001 | 0.068 | 0.001 |
| | 183.6 | -14.6 | 71 | 969.889 | 3.485 | 246.962 | 14.109 | 0.168 | 0.001 | 0.065 | 0.004 |
| | 209.79 | 33.29 | 11112 | 1005.956 | 8.371 | 268.209 | 2.376 | 0.169 | 0.002 | 0.07 | 0.001 |
| | 229.68 | 33.29 | 11112 | 989.396 | 7.392 | 269.625 | 4.130 | 0.166 | 0.002 | 0.069 | 0.001 |

| | | | | | | | | | | |
|---|---|---|---|---|---|---|---|---|---|---|
| 261.9 | -18.3 | *71 | 968.136 | 5.754 | 275.649 | 5.115 | 0.167 | 0.002 | 0.074 | 0.001 |
| 290.3 | -10.6 | *71 | 982.879 | 5.100 | 285.247 | 2.892 | 0.168 | 0.001 | 0.074 | 0.001 |
| 291.6 | -18.3 | *71 | 954.753 | 7.333 | 293.323 | 18.942 | 0.169 | 0.002 | 0.079 | 0.005 |
| 310.2 | 33.24 | 11112 | 986.362 | 6.670 | 270.331 | 2.831 | 0.169 | 0.002 | 0.071 | 0.001 |
| 310.49 | 33.24 | 11112 | 967.696 | 6.247 | 274.276 | 2.229 | 0.166 | 0.002 | 0.072 | 0.001 |
| 310.79 | 33.24 | 11112 | 977.626 | 6.432 | 276.445 | 2.488 | 0.168 | 0.002 | 0.072 | 0.001 |
| 329.6 | 33.25 | 11112 | 972.537 | 9.213 | 264.093 | 2.611 | 0.167 | 0.002 | 0.069 | 0.001 |
| 329.9 | 33.25 | 11112 | 969.428 | 7.788 | 269.921 | 2.416 | 0.166 | 0.002 | 0.071 | 0.001 |
| 330.19 | 33.25 | 11112 | 968.368 | 7.889 | 267.138 | 2.133 | 0.166 | 0.002 | 0.07 | 0.001 |

**Previously reported in Cartwright et al. (2015).*

0

**Table 4:** Mean IRAC and SpeX/LXD albedos

| | | | Prog. 11112 | | Prog. 71 | | SpeX/LXD | |
|---|---|---|---|---|---|---|---|---|
| Target | IRAC Channel | Hemisphere | Geo. Albedo | ΔGeo. Alb. (1σ) | Geo. Albedo | ΔGeo. Alb. (1σ) | Geo. Albedo | ΔGeo. Alb. (1σ) |
| Miranda | 1 | Leading | 0.194 | 0.013 | - | - | - | - |
| | 1 | Trailing | 0.194 | 0.014 | - | - | - | - |
| Ariel | 1 | Leading | 0.188 | 0.004 | 0.188 | 0.003 | 0.203 | 0.005 |
| | 1 | Trailing | 0.220 | 0.004 | 0.232 | 0.003 | - | - |
| | 2 | Leading | 0.083 | 0.004 | 0.092 | 0.005 | 0.229 | 0.007 |
| | 2 | Trailing | 0.112 | 0.005 | 0.124 | 0.005 | - | - |
| Umbriel | 1 | Leading | 0.151 | 0.002 | 0.153 | 0.002 | - | - |
| | 1 | Trailing | 0.154 | 0.001 | 0.158 | 0.001 | - | - |
| | 2 | Leading | 0.072 | 0.002 | 0.078 | 0.004 | - | - |
| | 2 | Trailing | 0.074 | 0.002 | 0.077 | 0.003 | - | - |
| Titania | 1 | Leading | 0.158 | 0.001 | 0.160 | 0.001 | 0.148 | 0.002 |
| | 1 | Trailing | 0.161 | 0.001 | 0.160 | 0.001 | 0.162 | 0.002 |
| | 2 | Leading | 0.061 | 0.000 | 0.063 | 0.002 | - | - |
| | 2 | Trailing | 0.066 | 0.000 | 0.066 | 0.001 | - | - |
| Oberon | 1 | Leading | 0.165 | 0.001 | 0.167 | 0.001 | 0.166 | 0.003 |
| | 1 | Trailing | 0.167 | 0.001 | 0.168 | 0.001 | 0.167 | 0.003 |
| | 2 | Leading | 0.067 | 0.001 | 0.067 | 0.001 | - | - |
| | 2 | Trailing | 0.071 | 0.001 | 0.076 | 0.002 | - | - |

**Table 5:** Ariel's five IRAC albedo zones

| | | | Prog. 11112 | | | Prog. 71 | | | IRAC (combined) | | | SpeX/LXD | |
|---|---|---|---|---|---|---|---|---|---|---|---|---|---|
| IRAC Channel | Albedo Region | Subsolar Long. Range (°) | Num. Data Points | Geo. Albedo | ΔGeo. Alb. (1$\sigma$) | Num. Data Points | Geo. Albedo | ΔGeo. Alb. (1$\sigma$) | Num. Data Points | Geo. Albedo | ΔGeo. Alb. (1$\sigma$) | Geo. Albedo | ΔGeo. Alb. (1$\sigma$) |
| 1 | UF | 3 - 8 | 4 | 0.201 | 0.003 | 0 | - | - | **4** | **0.201** | **0.003** | 0.203 | 0.005 |
| 1 | LE | 54 - 139 | 5 | 0.177 | 0.002 | 5 | 0.188 | 0.004 | **10** | **0.182** | **0.003** | - | - |
| 1 | AU | 188 | 1 | 0.206 | 0.005 | 0 | - | - | **1** | **0.206** | **0.005** | - | - |
| 1 | PT | 239 - 242 | 3 | 0.215 | 0.002 | 0 | - | - | **3** | **0.215** | **0.002** | - | - |
| 1 | CT | 254 - 277 | 3 | 0.229 | 0.005 | 2 | 0.232 | 0.004 | **5** | **0.230** | **0.003** | 0.229 | 0.007 |
| 2 | UF | - | 0 | - | - | 0 | - | - | **0** | **-** | **-** | - | - |
| 2 | LE | 54 - 139 | 5 | 0.083 | 0.004 | 5 | 0.093 | 0.007 | **10** | **0.088** | **0.004** | - | - |
| 2 | AU | 188 | 1 | 0.089 | 0.009 | 0 | - | - | **1** | **0.089** | **0.009** | - | - |
| 2 | PT | 239 - 242 | 3 | 0.108 | 0.004 | 0 | - | - | **3** | **0.108** | **0.004** | - | - |
| 2 | CT | 254 - 277 | 3 | 0.124 | 0.003 | 2 | 0.124 | 0.006 | **5** | **0.124** | **0.003** | - | - |

**Table 6:** *F*-test analysis of IRAC light curves

| Satellite | IRAC Albedo | *F*-test Ratio | Sample Size (n) | Mean Model Deg. Freedom (n - 1) | Sinusoidal Model Deg. Freedom (n - 3) | Probability (*p*) | Reject Null Hypothesis? |
|---|---|---|---|---|---|---|---|
| Ariel | Ch.1 | 135.08 | 23 | 22 | 20 | << 0.00001 | Yes |
| | Ch.2 | 124.54 | 19 | 18 | 16 | << 0.00001 | Yes |
| Umbriel | Ch.1 | 4.59 | 23 | 22 | 20 | $5.60 \times 10^{-4}$ | Yes |
| | Ch.2 | 5.09 | 23 | 22 | 20 | $2.71 \times 10^{-4}$ | Yes |
| Titania | Ch.1 | 2.88 | 24 | 23 | 21 | $8.83 \times 10^{-3}$ | Yes |
| | Ch.2 | 78.03 | 24 | 23 | 21 | << 0.00001 | Yes |
| Oberon | Ch.1 | 6.64 | 23 | 22 | 20 | $3.74 \times 10^{-5}$ | Yes |
| | Ch.2 | 29.22 | 23 | 22 | 20 | << 0.00001 | Yes |

**Table 7:** Best fit synthetic spectra for IRAC Ch.1 and Ch.2 albedos

| | Leading Hemisphere | | Trailing Hemisphere | | |
|---|---|---|---|---|---|
| Satellite | Model Components | Mix (%) | Model Components | Mix (%) | Reduced $\chi^2$ Statistic |
| Miranda | 10 µm $H_2O$ | 28.1 | 10 µm $H_2O$ | 28.1 | - |
| | 1 µm $H_2O$ | 66.9 | 1 µm $H_2O$ | 66.9 | |
| | 0.3 µm $H_2O$ | 2.2 | 0.3 µm $H_2O$ | 2.2 | |
| | 1 µm amorphous C | 2.8 | 1 µm amorphous C | 2.8 | |
| Ariel | 2 µm $H_2O$ | 15.6 | 2 µm $H_2O$ | 55.4 | 0.212 (Lead.) |
| | 1 µm $H_2O$ | 79.6 | 1 µm $H_2O$ | 41.5 | 0.432 (Trail.) |
| | 0.2 µm $H_2O$ | 1.8 | 0.2 µm $H_2O$ | 0.9 | |
| | 1 µm amorphous C | 3.0 | 1 µm amorphous C | 2.2 | |
| Umbriel | 10 µm H2O | 10.8 | 10 µm $H_2O$ | 6.0 | - |
| | 2 µm $H_2O$ | 12.5 | 2 µm $H_2O$ | 16.1 | |
| | 1 µm $H_2O$ | 40.5 | 1 µm $H_2O$ | 41.2 | |
| | 0.3 µm $H_2O$ | 4.2 | 0.3 µm $H_2O$ | 4.2 | |
| | 10 µm amorphous C | 32.1 | 10 µm amorphous C | 32.6 | |
| Titania | 10 µm $H_2O$ | 20.4 | 2 µm $H_2O$ | 11.0 | 0.109 (Lead.) |
| | 1 µm $H_2O$ | 58.6 | 1 µm $H_2O$ | 71.1 | 0.098 (Trail.) |
| | 0.3 µm $H_2O$ | 3.6 | 0.3 µm $H_2O$ | 5.1 | |
| | 10 µm amorphous C | 10.1 | 1 µm amorphous C | 7.8 | |
| | 10 µm Triton tholin | 7.4 | 10 µm Triton tholin | 5.0 | |
| Oberon | 10 µm $H_2O$ | 21.1 | 10 µm $H_2O$ | 3.8 | 0.190 (Lead.) |
| | 2 µm $H_2O$ | 1.0 | 1 µm $H_2O$ | 71.0 | 0.146 (Trail.) |
| | 1 µm $H_2O$ | 58.0 | 0.3 µm $H_2O$ | 4.2 | |
| | 0.3 µm $H_2O$ | 1.8 | 10 µm amorphous C | 17.1 | |
| | 10 µm amorphous C | 14.1 | 10 µm Triton tholin | 4.0 | |
| | 10 µm Triton tholin | 4.0 | | | |

*All synthetic spectra shown in Figure 6.*

**Table 8:** Best fit synthetic spectra for Ariel albedo zones (no $CO_2$ models)

| Ariel Albedo Zone | Model Components | Mix (%) | Reduced $\chi^2$ Statistic |
|---|---|---|---|
| UF | 2 µm $H_2O$ | 35.1 | 0.379 |
| | 1 µm $H_2O$ | 60.6 | |
| | 0.2 µm $H_2O$ | 1.3 | |
| | 1 µm amorphous C | 3.0 | |
| | 1 µm amorphous C | 3.0 | |
| LE | 10 µm H2O | 6.7 | - |
| | 2 µm $H_2O$ | 29.9 | |
| | 1 µm $H_2O$ | 58.5 | |
| | 0.2 µm $H_2O$ | 1.9 | |
| | 1 µm amorphous C | 3.0 | |
| AU | 2 µm $H_2O$ | 20.1 | - |
| | 1 µm $H_2O$ | 76.6 | |
| | 0.2 µm $H_2O$ | 1.1 | |
| | 1 µm amorphous C | 2.3 | |
| PT | 2 µm $H_2O$ | 51.5 | - |
| | 1 µm $H_2O$ | 45.0 | |
| | 0.2 µm $H_2O$ | 1.0 | |
| | 1 µm amorphous C | 2.5 | |
| *CT | 2 µm $H_2O$ | 67.0 | 0.409 |
| | 1 µm $H_2O$ | 31.0 | |
| | 0.2 µm $H_2O$ | 0.8 | |
| | 1 µm amorphous C | 1.2 | |

**Ariel CT model shown in Figure 7c.*

**Table 9:** Best fit synthetic spectra for Ariel's CT albedo zone ($CO_2$-included models)

| $CO_2$ Mixing Regime | Model Components | Mix (%) | Reduced $\chi^2$ Statistic |
|---|---|---|---|
| | 2 µm $H_2O$ | 51.3 | 0.438 |
| | 1 µm $H_2O$ | 25.2 | |
| | 0.2 µm $H_2O$ | 1.1 | |
| | 1 µm amorphous C | 3.4 | |
| Intimate | 1 µm $CO_2$ ice | 19.0 | |
| | 2 µm $H_2O$ | 47.1 | 0.426 |
| | 1 µm $H_2O$ | 22.3 | |
| | 0.2 µm $H_2O$ | 0.6 | |
| | 1 µm amorphous C | 1.0 | |
| Intimate | 10 µm $CO_2$ ice | 27.0 | |
| | 50 µm $H_2O$ | 36.5 | 1.066 |
| | 10 µm $H_2O$ | 34.4 | |
| | 0.3 µm $H_2O$ | 0.6 | |
| | 12.5 µm amorphous C | 1.6 | |
| Areal | 50 µm $CO_2$ ice | 5.4 | |
| | 10 µm $CO_2$ ice | 21.6 | |
| | 10 µm $H_2O$ | 65.3 | 0.528 |
| | 1 µm $H_2O$ | 19.1 | |
| | 0.2 µm $H_2O$ | 0.9 | |
| | 1 µm amorphous C | 1.9 | |
| Areal | 50 µm $CO_2$ ice | 2.6 | |
| | 10 µm $CO_2$ ice | 10.4 | |
| | 10 µm $H_2O$ | 64.6 | 0.579 |
| | 1 µm $H_2O$ | 19.0 | |
| | 0.2 µm $H_2O$ | 0.9 | |
| | 1 µm amorphous C | 1.7 | |
| Areal | 1 µm $CO_2$ ice | 13.9 | |

*Synthetic spectra shown in Figure 7a and 7b.*

**Table 10:** Example $H_2O$ ice synthetic spectra compared to Ariel CT albedo zone

| Model Components | Mix (%) | Reduced $\chi^2$ Statistic | Modeled Ch.1 Albedo | Modeled Ch.2 Albedo |
|---|---|---|---|---|
| *1 µm $H_2O$ ice | 100 | 0.453 | 0.240 | 0.089 |
| *10 µm $H_2O$ ice | 100 | 0.458 | 0.241 | 0.236 |
| *100 µm $H_2O$ ice | 100 | 1.476 | 0.112 | 0.099 |
| 1 µm $H_2O$ ice<br>10 µm $H_2O$ ice | 50<br>50 | 0.430 | 0.240 | 0.097 |
| 10 µm $H_2O$ ice<br>100 µm $H_2O$ ice | 50<br>50 | 0.418 | 0.222 | 0.216 |

**Shown in Figure 7c.*

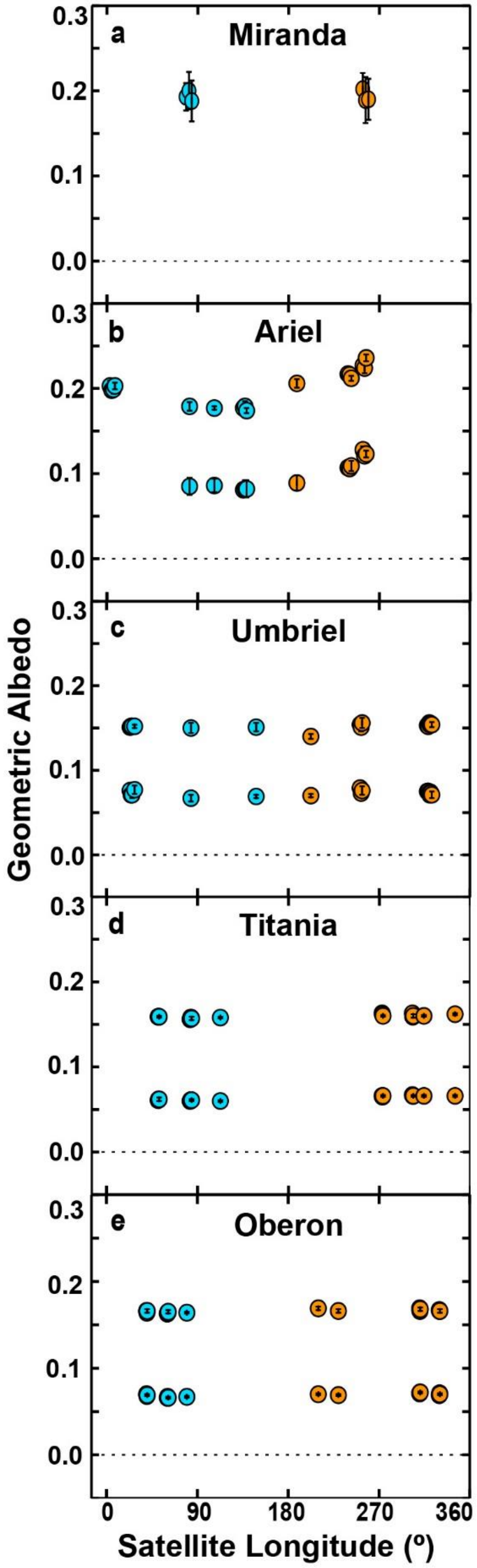

**Figure 1:** Program 11112 Spitzer/IRAC geometric albedos and 1$\sigma$ uncertainties for (a) Miranda, (b) Ariel, (c) Umbriel, (d) Titania, and (e) Oberon, plotted as a function of satellite longitude. Ch.1 (top rows) and Ch.2 (bottom rows) albedos are shown for the leading (blue) and trailing (orange) hemisphere of each moon.

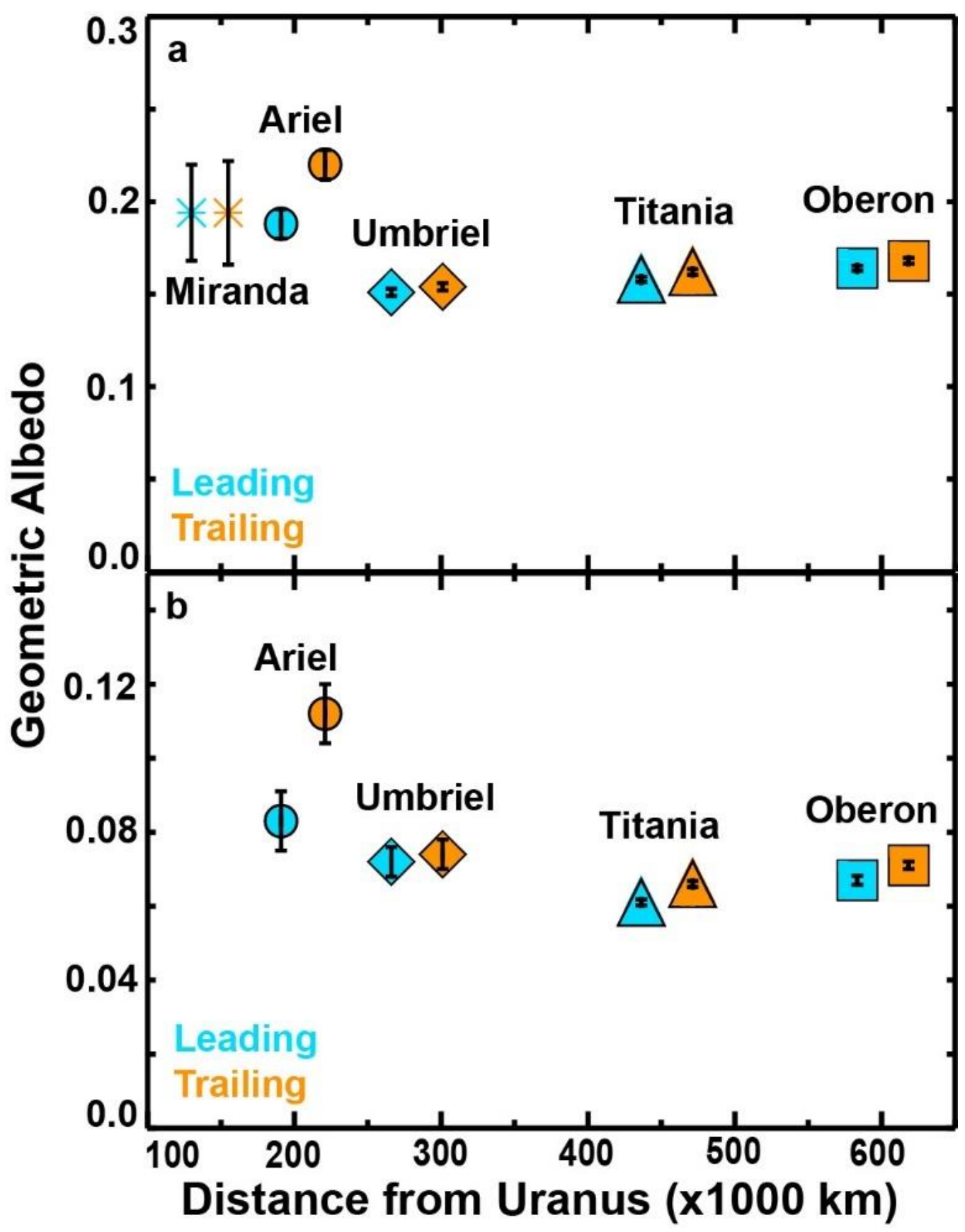

**Figure 2:** Program 11112 IRAC (a) Ch.1 and (b) Ch.2 mean leading (blue) and trailing (orange) geometric albedos and 2$\sigma$ uncertainties for each satellite. In both plots, each moon is represented by asterisks (Miranda), circles (Ariel), diamonds (Umbriel), triangles (Titania), and squares (Oberon).

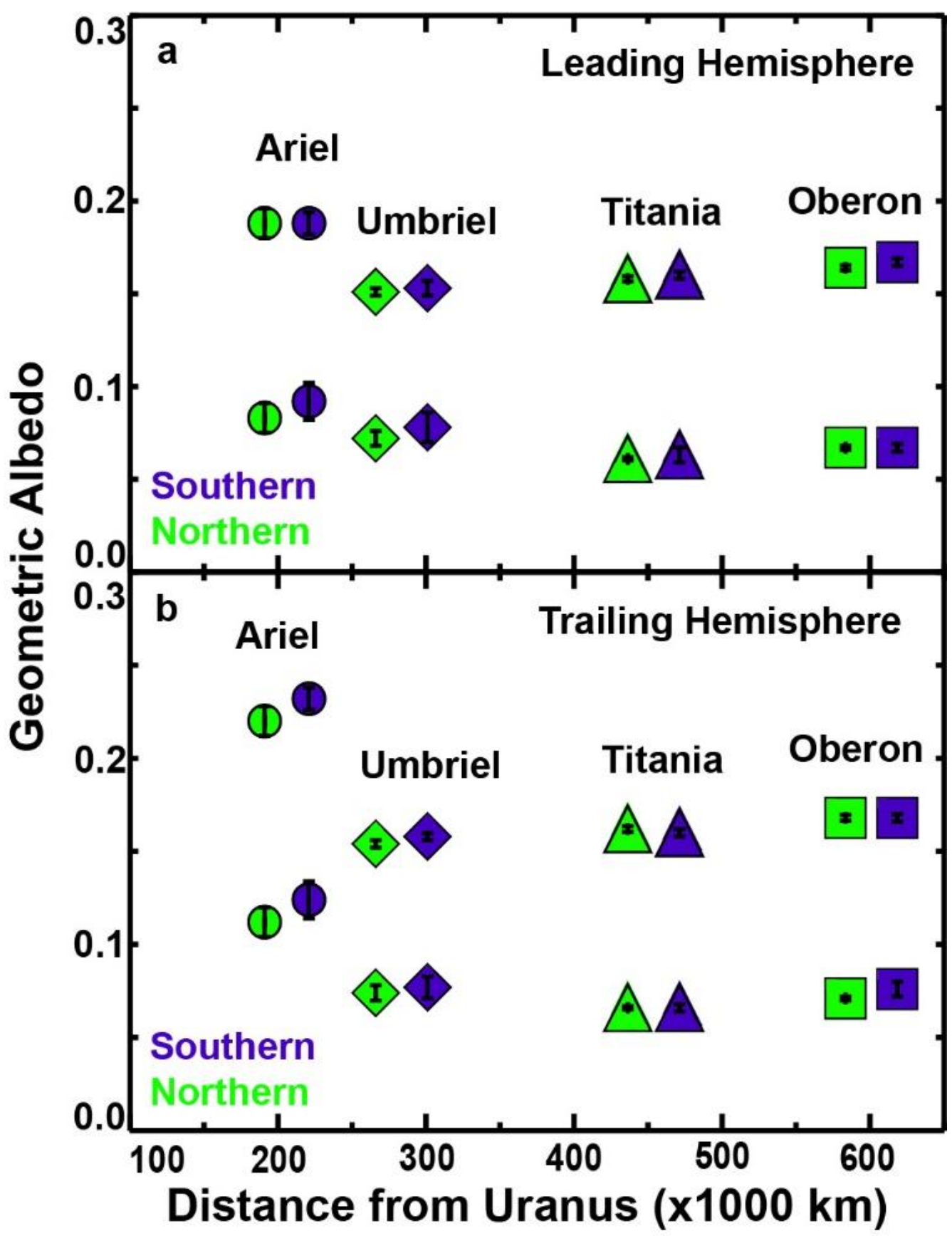

**Figure 3:** Program 71 (southern, purple) and 11112 (northern, green) IRAC mean geometric albedos and 2$\sigma$ uncertainties for the (a) leading and (b) trailing hemisphere of each satellite. In both plots, each moon is represented by circles (Ariel), diamonds (Umbriel), triangles (Titania), and squares (Oberon).

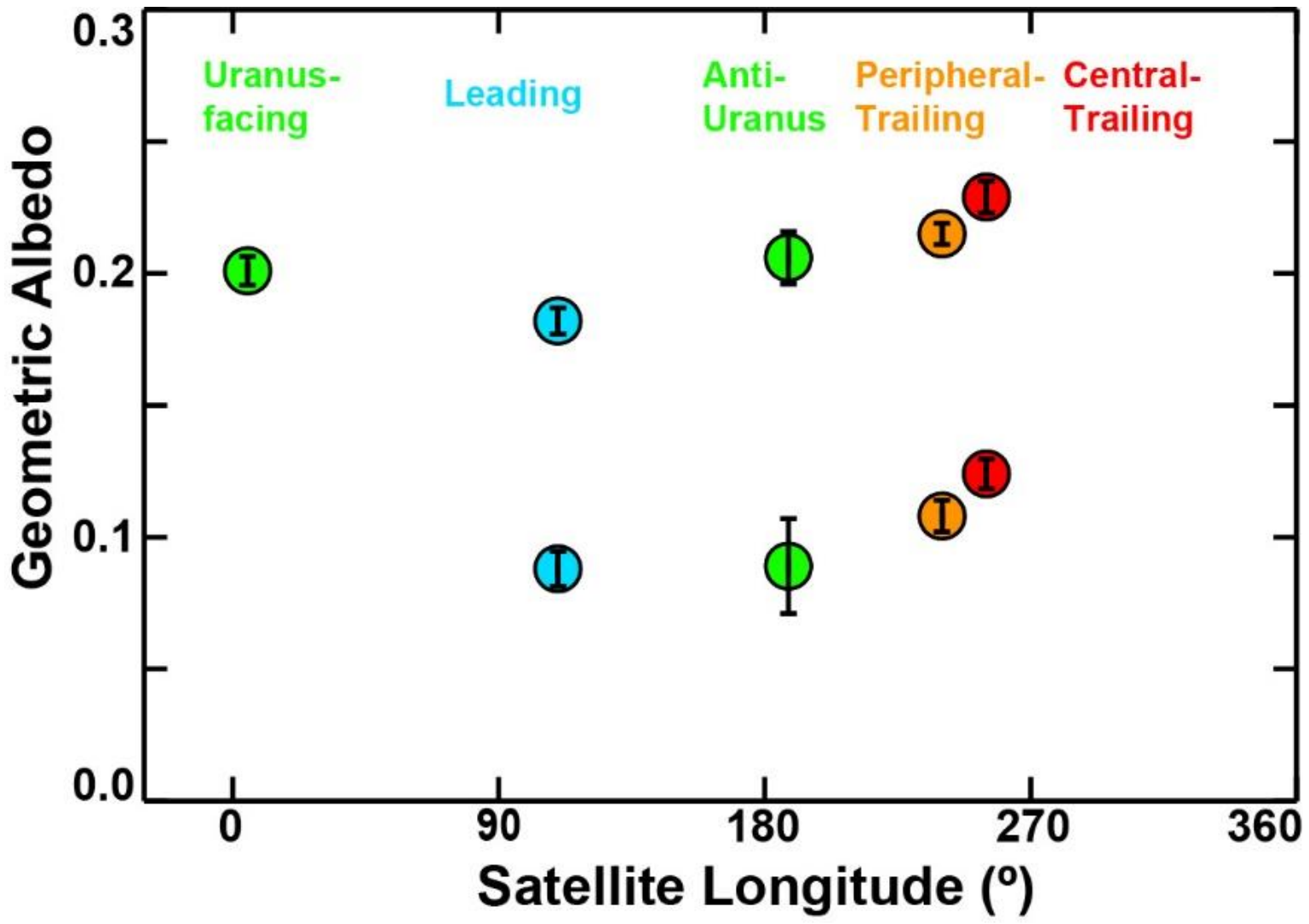

**Figure 4:** Mean IRAC Ch.1 (top row) and Ch.2 (bottom row) geometric albedos and $2\sigma$ uncertainties for each of the albedo zones identified on Ariel: Uranus-facing (UF) and Anti-Uranus (AU) (green), Leading (LE) (blue), Peripheral-Trailing (PT) (orange), and Central-Trailing (CT) (red).

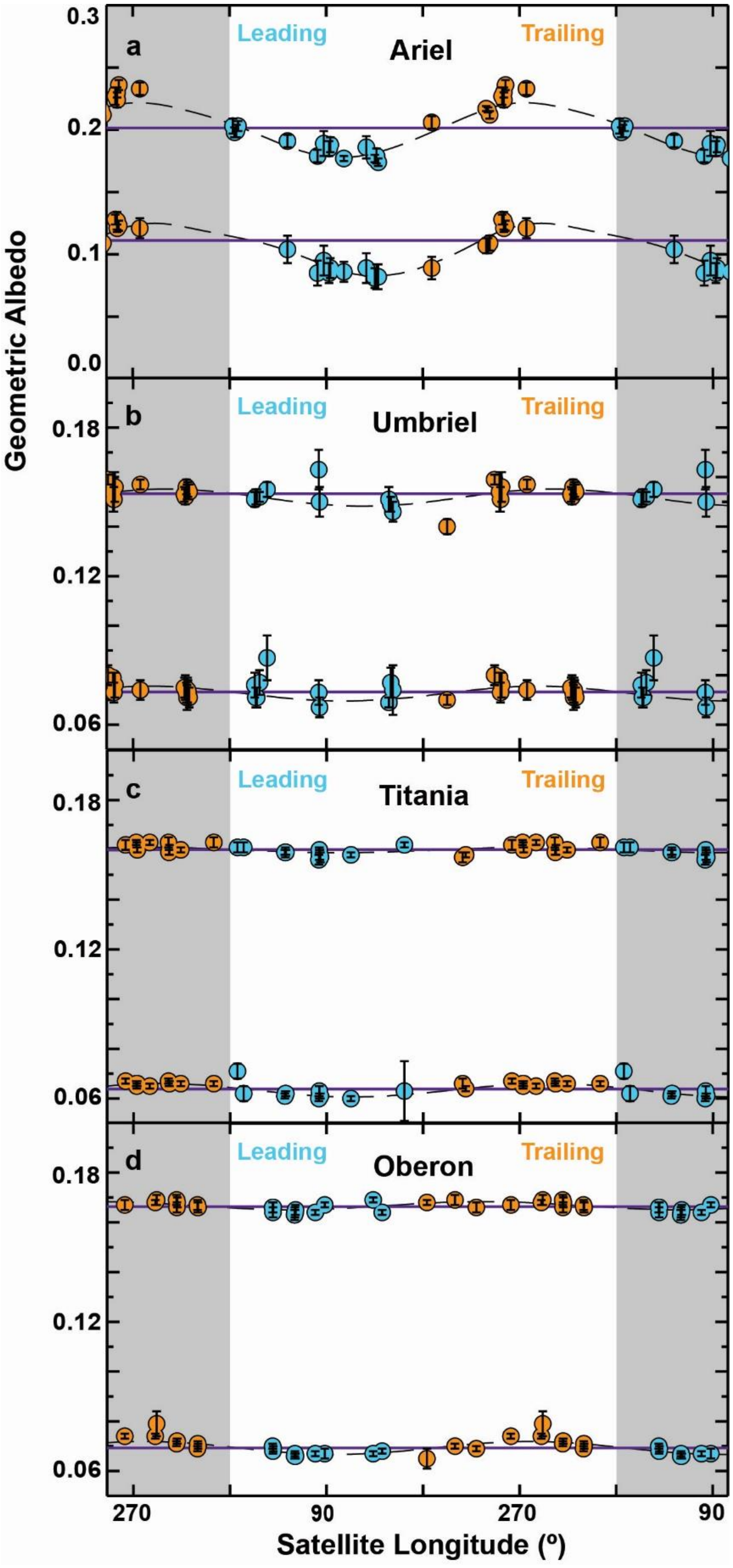

**Figure 5:** Program 71 and 11112 Spitzer/IRAC geometric albedos and 1$\sigma$ uncertainties for (a) Ariel, (b) Umbriel, (c) Titania, and (d) Oberon plotted as a function of satellite longitude. Ch.1 (top rows) and Ch.2 (bottom rows) albedos are shown for the leading (blue) and trailing (orange) hemisphere of each moon. Dashed lines represent sinusoidal fits to the data and the solid purple lines show the mean Ch.1 and Ch.2 albedos for each moon. Duplicate longitudes are shown to highlight periodic trends in the IRAC albedos for each moon (gray-toned regions). The maxima of these sinusoidal models are free parameters and are not locked to a specific longitude. The y-axis of the Umbriel, Titania, and Oberon plots are scaled to 0.05 to 0.2 geometric albedo, whereas the y-axis of the Ariel plot ranges from 0.0 to 0.3 geometric albedo.

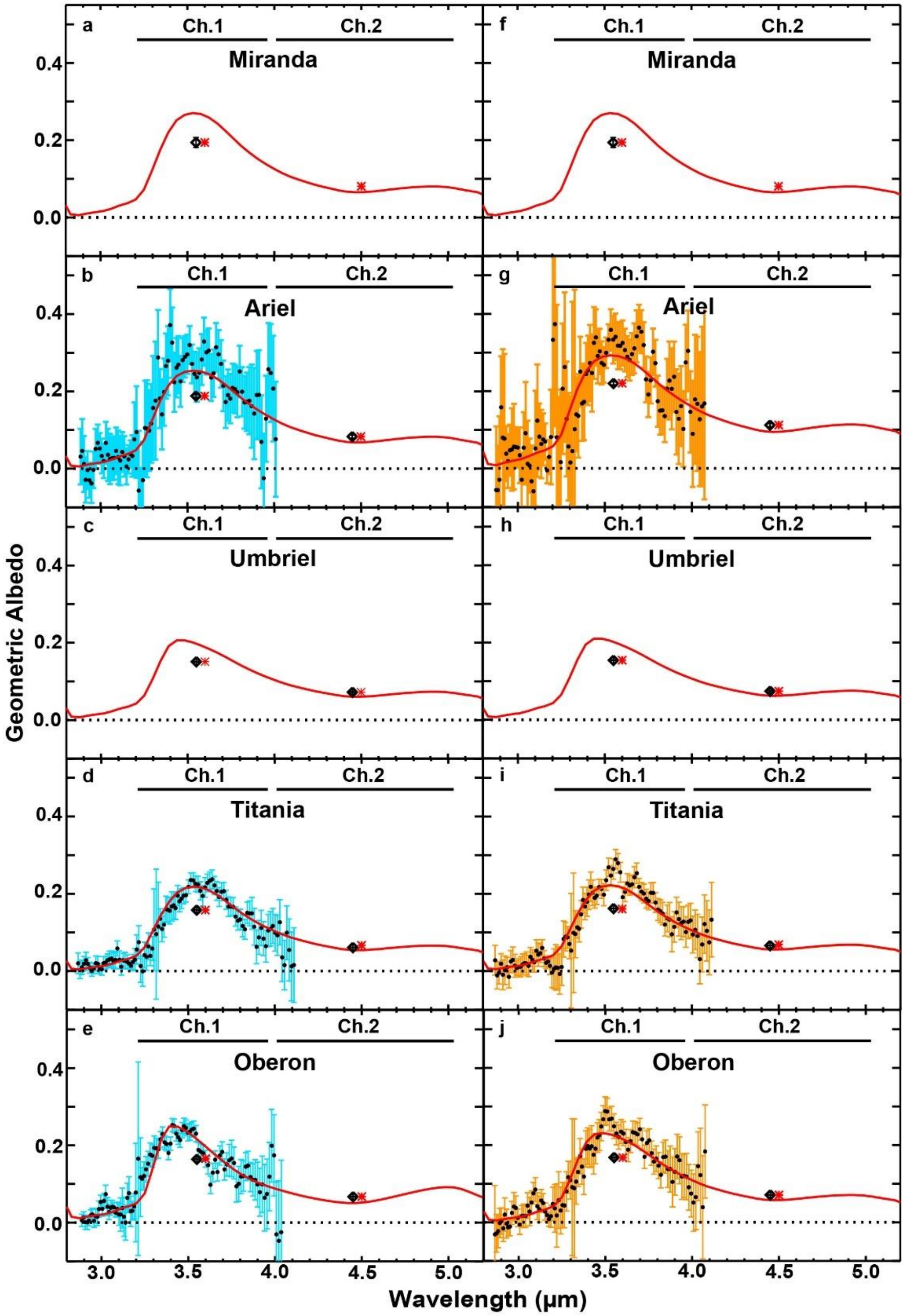

**Figure 6:** Best fit synthetic spectra (red lines) for the mean IRAC Ch.1 and Ch.2 albedos for each moon's leading (a-e) and trailing (f-j) hemisphere. The modeled Ch.1 and Ch.2 albedos (red asterisks) for these synthetic spectra are shown to the right of the measured albedos (black diamonds). FWHM bandwidth for Ch.1 and Ch.2 (black bars) are indicated. Published SpeX/LXD spectra for the leading (blue) and trailing (orange) hemisphere of Ariel, Titania, and Oberon are also shown (originally presented in Cartwright et al., 2018). Synthetic spectra details are summarized in Table 7.

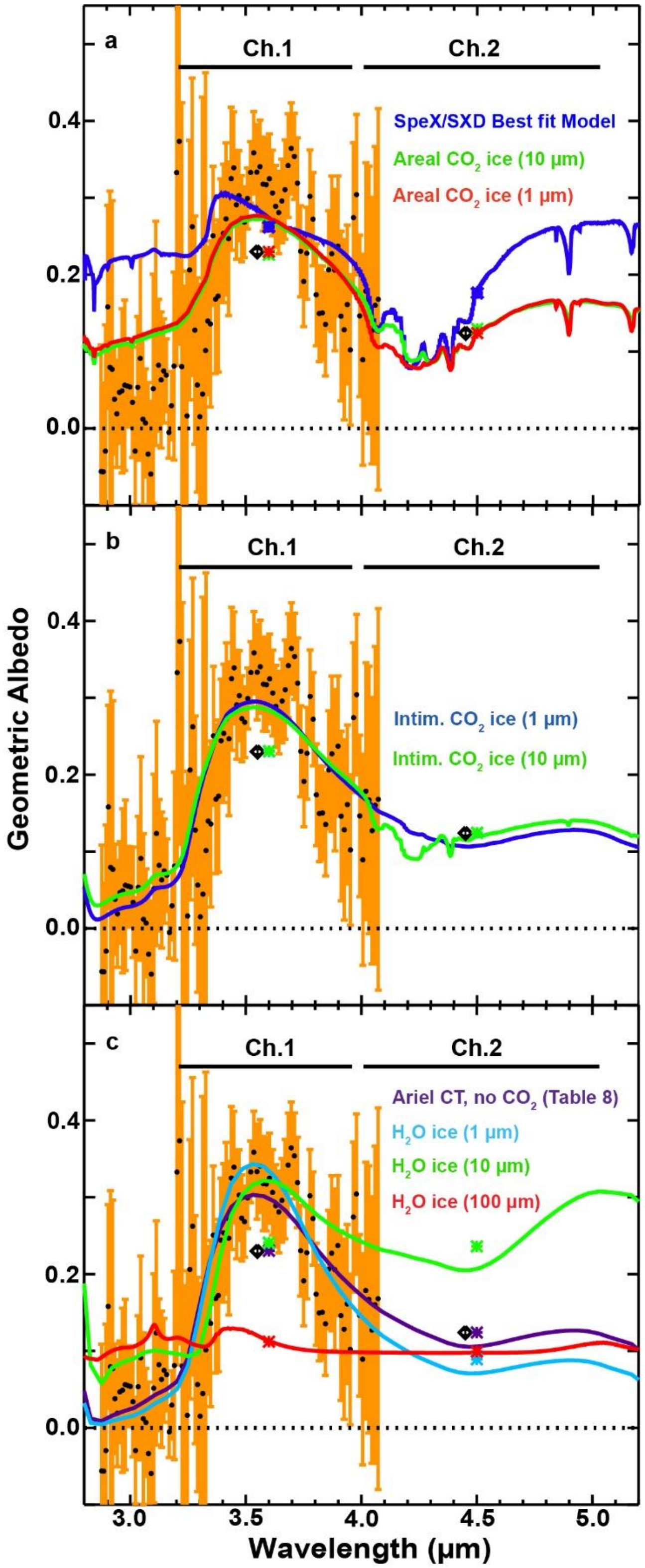

**Figure 7:** Examples of synthetic spectra fit to Ariel's Central-Trailing albedo zone: (a) models that include areally mixed $CO_2$ ice, (b) models with intimately mixed $CO_2$ ice, (c) models without $CO_2$ ice. The synthetic spectra, their modeled Ch.1 and Ch.2 albedos, and their descriptive labels all share the same color in each plot. The measured Ch.1 and Ch.2 albedos for the Central-Trailing albedo region are also shown (black diamonds). FWHM bandwidths for Ch.1 and Ch.2 (black bars) are indicated. Published SpeX/LXD spectrum for the trailing hemisphere of Ariel is also shown (originally presented in Cartwright et al., 2018).